\documentclass[%
 reprint,
 amsmath,amssymb,
 aps,
]{revtex4-2}

\usepackage{graphicx}
\usepackage{dcolumn}
\usepackage{bm}
\usepackage{hyperref}

\usepackage{cleveref}
\usepackage{color}
\usepackage{mhchem}

\usepackage[font=small,labelfont=bf]{caption}
\usepackage{subcaption}

\usepackage{float}
\usepackage{placeins}
\usepackage{tikz}
\usetikzlibrary{positioning, matrix, shapes.geometric}
\usepackage{algorithm}
\usepackage{algorithmic}
\usepackage{comment}
\usepackage{amssymb}
\usepackage{amsmath}

\usepackage{makecell}
\usepackage{tabularx}
\newcommand{\orcid}[1]{\href{https://orcid.org/#1}{\includegraphics[height=\fontcharht\font`\B]{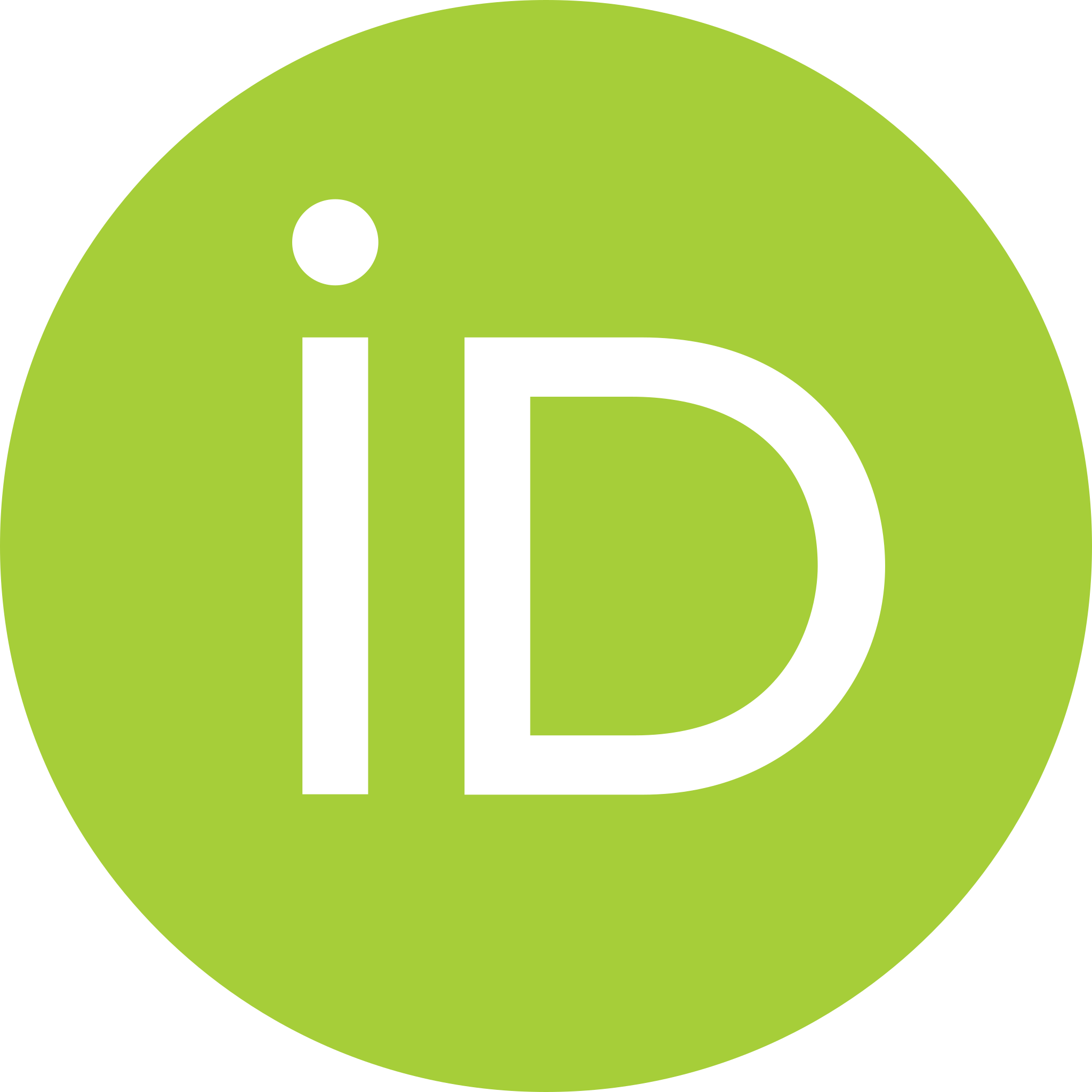}}}

\begin{document}


\title{Feldman-Cousins' ML Cousin: Sterile Neutrino Global Fits using Simulation-Based Inference}
\author{Joshua Villarreal \orcid{0000-0001-9690-1310}}
\email{villaj@mit.edu}
\author{John Hardin \orcid{0000-0001-8871-8065}}%
\author{Janet M. Conrad \orcid{0000-0002-6393-0438}}
\affiliation{%
 Massachusetts Institute of Technology \\
 77 Massachusetts Avenue \\
 Cambridge, MA 02139, USA
}%

\date{\today}

\begin{abstract}
For many small-signal particle physics analyses, Wilks' theorem, a simplifying assumption that presumes log-likelihood asymptotic normality, does not hold. The most common alternative approach applied in particle physics is a highly computationally expensive procedure put forward by Feldman and Cousins. 
When many experiments are combined for a global fit to data, deviations from Wilks' theorem are exacerbated, and Feldman-Cousins becomes computationally intractable. We present a novel, machine learning-based procedure that can approximate a full-fledged Bayesian analysis 200 times faster than the Feldman-Cousins method. We demonstrate the utility of this novel method by performing a joint analysis of electron neutrino/antineutrino disappearance data within a single sterile neutrino oscillation framework. Although we present a prototypical simulation-based inference method for a sterile neutrino global fit, we anticipate that similar procedures will be useful for global fits of all kinds, especially those in which Feldman-Cousins is too computationally expensive to use.
\end{abstract}


\maketitle

\begin{table*}[htb!]
\centering
\resizebox{\textwidth}{!}{
\begin{tabular}{|l|c|c|c|c|c|}  \hline
Name & Ref. &  L [m] & E [MeV] & Significance & Short Description \\  \hline \hline
\multicolumn{6}{|c|}{Reactor Experiments ($\bar \nu_e$)}  \\ \hline

STEREO & \cite{Almazan2023} & 9-11 & 1.625-7.125 &  NR & ILL research reactor, extended detector\\

PROSPECT & \cite{prospectResults} &  7-10 & 0.8-7.2 &  NR & ORNL research reactor, extended detector\\

DANSS & \cite{Skrobova2023} & 10-12 & 1-7 &  $2.3 \sigma$ & Kalinin power reactor, moving baseline system\\

NEOS/RENO & \cite{PhysRevD.105.L111101} & 24/294 & 1-10 & NR & \makecell{Hanbit power reactor, two detectors, with \\ RENO near detector used as SM flux baseline} \\ \hline
\multicolumn{6}{|c|}{Source Experiments ($\nu_e$)} \\ \hline
SAGE \& GALLEX & \cite{Gavrin2013,KOSTENSALO2019542,Giunti:2010zu}, & 0-0.7 \& 0-1.9 & \makecell{0.426, 0.747, 0752,\\ 0.811, 0.813} & $2.3 \sigma$ & $^{51}$Cr and $^{37}$Ar sources; single segment $^{71}$Ga target \\
BEST & \cite{PhysRevLett.128.232501, PhysRevD.105.L051703} & 0-1.7 & 0.426, 0.747, 0752 & $>5 \sigma$& $^{51}$Cr source; two-segment $^{71}$Ga target \\ \hline
\end{tabular}}
\caption {\label{tab:experiments}Relevant parameters of experiments used in this study.
Characteristic $L$ and $E$ values are approximate, see reference (column 2) for more details.  Column 5 provides the reported significance of a 3+1 signal in numbers of standard deviations (or NR if the best fit point is not reported).}
\end{table*}

\section{\label{sec:introduction}Introduction}

Searching for rare particles or physics processes often involves synthesizing data from multiple experiments and signatures. For example, the first estimates of the top quark mass combined data from numerous electroweak experiments \cite{ALEPH:1995ac}; the Higgs boson search drew on a wide range of experimental data sets \cite{PDG2012Higgs}; and the establishment of the three-neutrino oscillation model relied on diverse neutrino sources and detectors \cite{Gonzalez-Garcia:2012hef}. These exercises in \textit{global fitting} were instrumental in achieving the 1999, 2013, and 2015 Nobel Prizes. In addition to validating and excluding theoretical predictions, global fits have also been valuable for highlighting unexplored model parameter spaces for future studies, honing the direction of particle physics experiments.

In a traditional frequentist global fit, a test statistic is computed over a model's parameter space and profiled to assess the stability and significance of a fit by building a \textit{confidence level} (CL). The simplest approach involves calculating the model's likelihood across the parameter space, profiling the log-likelihood difference relative to a null (no-signal) hypothesis, and describing the test statistic with a $\chi^2$ distribution, as per Wilks' theorem. This method is efficient for deriving CLs, but issues like varying degrees of freedom and non-Gaussian systematic uncetainties can misrepresent the CLs, failing frequentist consistency checks in many mock experimental trials \cite{Hardin:2022qdh}.

To address this, the Feldman-Cousins method, which compares data from simulated trials across the entire parameter space to the null, is preferred \cite{feldman-cousins}. However, this approach is highly computationally expensive, requiring extensive trials in a fine grid across the available parameter space and repeated evaluation of the test statistic. Thus, despite the Feldman-Cousins approach being widely regarded as the best choice for correctly characterizing fit results, many analyses are forced to resort to Feldman-Cousins ``spot checks'' in a more limited region of parameter space to quantify the deviation from the Wilks' approach rather than complete a high-fidelity data analysis.   
An alternative approach that is faster to evaluate is the CL$_s$ \footnote{Not to be confused with confidence levels, CLs.} method\cite{Qian:2014nha}; however, this is only applicable for setting limits on model parameters, and not for identifying allowed regions, making it less relevant to the goals of most signal-searching global fits.

The frequentest approaches described above have a Bayesian alternative. Using Bayes' rule, one establishes a \textit{prior distribution} over fit parameters, uses realized experimental data to build a \textit{likelihood}, and computes a \textit{posterior distribution} from which \textit{credible regions} (CRs) can be computed. Depending on the problem's complexity, computing likelihoods (using methods like Markov Chain Monte Carlo) may be computationally expensive or even intractable. 



Neither frequentist nor Bayesian approaches are immune from heavy computational overhead. To this end, enhancement or even replacement of these traditional fitting procedures with \textit{likelihood-free} or \textit{simulation-based inference} (SBI) has been a major recent effort in scientific computing.  The aim of SBI is to develop statistical or machine learning (ML) techniques that can efficiently learn the likelihood function, posterior distribution, or some other quantity (like the likelihood ratio) used for interpreting the results of a fit. Recent advances in deep density estimation via normalizing flows, the primary tooling for this work, have illuminated the huge potential of SBI for efficient data analysis, particularly in the context of particle physics experiments. For a review of SBI and its applications, see Ref.~\cite{Cranmer2020}; for a community-led project compilation of SBI methodological developments and novel domain-specific applications, see Refs.~\cite{sbidotorg, neuralsbigh}.

This paper explores how to extend SBI methods from single-experiment analyses to global fits involving arbitrary combinations of experiments. This approach to global fitting has multiple benefits:
\begin{itemize}
\item The ML algorithm is trained using trials, capturing information on the stochastic behavior of the data set expected at each parameter point.  
\item The data can be generated with systematic uncertainties that are not Gaussian, a point rarely incorporated into fits to single experiments or global fits.  
\item Analysis of a single experiment is very fast relative to traditional fits, and expanding from a single experiment to a global fit involves little increase in the total run time.
\end{itemize}

\subsection{Sterile Neutrino Searches via Electron-Neutrino Disappearance}

To demonstrate the viability and speed of this new global fitting technique compared to high-fidelity analysis frameworks like Feldman-Cousins, we choose the domain of searches for \textit{sterile neutrinos}---hypothetical noninteracting partners to the three known neutrino flavors in the Standard Model (SM). These searches rely on the quantum mechanical phenomenon of \textit{neutrino oscillations}, where one type of neutrino converts into another, a well-established behavior for SM neutrinos. Some short baseline (SBL) experiments have observed effects suggesting additional oscillation frequencies beyond SM predictions, while others have found no such evidence, casting doubt on the sterile neutrino hypothesis \cite{Hardin:2022muu}.

To resolve these conflicting results, we must analyze the SBL data as a whole through global fits within a consistent sterile neutrino model. The simplest such model (called ``3+1'') involves oscillations between three SM neutrinos and one sterile neutrino. However, fitting the 3+1 model is challenging: Wilks' approach is too simplistic \cite{Hardin:2022qdh}, and Feldman-Cousins is computationally prohibitive. SBI techniques naturally lend themselves useful towards analyzing experiments' preferences for or against a 3+1 sterile neutrino model. 

To demonstrate SBI's utility for informing sterile neutrino global fits, we will focus on fitting a subset of the SBL data, specifically from searches for the disappearance of electron-flavor neutrino interactions. This subset is ideal because it encompasses results from two families of experiments, featuring both anomalies and SM-consistent findings, making it a representative sample of the challenges faced by global fitters. Moreover, the analysis is simplified by the subset's reliance on just two model parameters. A comprehensive study involving all 3+1 oscillation channels through both traditional fitting frameworks and SBI techniques will be addressed in a future paper.

We can split electron-flavor disappearance experiments into two categories: those performed using antineutrinos produced at reactors and those using neutrinos emitted from highly radioactive nuclear decays.  In both cases the 3+1 model predicts the survival probability:
\begin{equation}
P_{\nu_e \rightarrow \nu_e} = 1-4|U_{e4}|^2 (1-|U_{e4}|^2) \sin^2(\Delta_{41}L/E),~ \label{PUe4}
\end{equation} 
where $\Delta_{41} \equiv 1.27 \Delta m_{41}^2$. This probability depends on the ratio of two experimental parameters:  $L$, the distance from the neutrino creation to the observation point in meters (sometimes called the \textit{baseline}), and $E$, the energy of the neutrino in MeV. The fit extracts two model parameters:  $|U_{e4}|$, the amplitude of the oscillation, and $\Delta m_{41}^2\,[\text{eV}^2]$, the oscillation frequency in $L/E$-space \footnote{Note that to keep $P_{\nu_e \to \nu_e} \in [0, 1]$, we require $|U_{e4}| \in [0, 1]$. For fits to electron disappearance experiments only, having $|U_{e4}| \leq 1 / \sqrt{2}$ still fully explores the model parameter space.}.

Table~\ref{tab:experiments} summarizes the experiments used in our mock global fit. The modern reactor experiments with the most extensive reach in 3+1 parameter space are STEREO, PROSPECT, DANSS, and NEOS. These reactors produce anti-electron neutrinos ($\bar{\nu}_e$) detected via inverse beta decay interactions $\overline{\nu}_e (p, n) e^+$, allowing precise reconstruction of baseline $L$ and antineutrino energy $E$. To detect sterile neutrino oscillations, the measured event rate as a function of $L/E$ is compared to SM expectations, though reactor flux predictions are known to be inaccurate \cite{Giunti:2021kab}. We address this by removing the overall normalization mismatch to perform ``shape-only" fits, reducing sensitivity to high values of $\Delta m_{41}^2$ ($\gtrapprox \mathcal{O} (10\,\text{eV}^2) $) where rapid oscillations averaged over many detector segments could mimic a normalization effect.

In contrast, SAGE, GALLEX and BEST use fluxes produced by radioactive sources with $\text{MCi}$-level decay rates. These source experiments employ monoenergetic electron neutrinos ($\nu_e$) produced through electron capture in $^{51}$Cr and $^{37}$Ar sources, paired with gallium detectors. The $\nu_e$ interactions convert gallium to germanium ($\nu_e (\ce{^{71}Ga}, \ce{^{71}Ge}) e^-$), which is periodically removed and counted to compare against well-predicted rates \cite{Elliott:2023xkb}. While low normalization uncertainty allows high sensitivity to non-zero $|U_{e4}|$, poor segmentation limits $\Delta m_{41}^2$ sensitivity, restricting measurements to $\Delta m_{41}^2 \gtrapprox \mathcal{O} (10\,\text{eV}^2)$. Thus, the reactor and source data sets are complementary.   

\subsection{Relevant Machine Learning Tooling}

\subsubsection{\label{subsec:nn-uncertainties}Neural Networks and Prediction Uncertainties}

Most neural networks (NNs) are limited to making deterministic predictions. However, for global fits, it's crucial that our networks can quantify predictive uncertainties meaningfully. Bayesian neural networks (BNNs) \cite{Neal:bayesianNN, MacKay:bayesianNN} address this challenge by placing distributions on the nodes in the hidden layers, enabling inference samples to be drawn from a learned probability distribution. Although theoretically advantageous for our problem, BNNs are practically infeasible due to the significant computational overhead required for training, similar to other nonparametric models like Gaussian processes.

Gal and Ghahramani \cite{pmlr-v48-gal16} propose an approximation to Bayesian techniques using dropout, a common neural network regularization method. In dropout, network weights are randomly ``dropped" (zeroed out) with a probability known as the \textit{dropout rate} at each training step. This process effectively trains an ensemble of smaller sub-networks, mitigating overfitting and underfitting when dropout is turned off during inference, as each node contributes to the final prediction \cite{JMLR:v15:srivastava14a}. However, if dropout is kept on during inference, as in Ref.~\cite{pmlr-v48-gal16}, the network becomes stochastic, with a randomly selected sub-network making each prediction. A key result from Gal and Ghahramani's work is that training a dropout-at-inference NN is procedurally similar to minimizing the Kullback-Leibler divergence in training a deep Gaussian process. Consequently, moments computed from a set of inference samples can be translated to prediction uncertainties,  a critical component of our SBI procedure described in Sec.~\ref{subsec:fcmlc-overview}.

\subsubsection{\label{normalizingflows}Normalizing Flows}

Normalizing flows (NFs) fall into a class of neural networks that learn a set of invertible transformations of probability distributions to sample from and estimate the density of a complex posterior distribution. NFs for density estimation were popularized in Ref.~\cite{pmlr-v37-rezende15}, and the notation for the operational principle of NFs from this reference is used herein.

NFs are based on the change-of-variables formula for probability distributions. That is, for $\mathbf{f}$ an invertible mapping of some $\mathbf{z}$ to $\mathbf{z}'$, if $\mathbf{z}$ is described by a distribution $q(\mathbf{z})$, then $\mathbf{z'}$ can be described by a distribution ${q(\mathbf{z}') = q (\mathbf{z}) | \det \partial \mathbf{f} / \partial \mathbf{z} |^{-1}}$. An arbitrarily complex distribution can thus be modeled as many such compositions of invertible mappings of a latent variable with some well-defined (e.g. Gaussian) base distribution.
The invertibility of the compositions described here are especially useful for gradient-based loss-minimization procedures in the backpropagation steps during NF training.

The customization of a normalizing flow to a variational inference task comes in part from the class(es) of invertible mappings used to estimate some posterior distribution. Through some trial-and-error, we have settled on a neural spline flow to handle the complex density estimation task relevant to this problem; in particular, the coupling layers of the normalizing flow (Fig.~\ref{fig:nf-architecture}) seek to learn parameters for a rational quadratic spline which relates the chained transformations \cite{NEURIPS2019_7ac71d43}.

\section{\label{sec:methodology}Methodology}

\begin{figure*}[t!]
\centering
\resizebox{\linewidth}{!}{
\begin{tikzpicture}[node distance=1.cm, every node/.style={rectangle, minimum width=.5cm, minimum height=0.5cm, draw, fill=blue!20, align=center}]

\node[draw, fill=gray!30] (input) {Input $\mathbf{x}_i$\\(Experimental realization)};
\node[draw, fill=white, below of=input, yshift=-0.5cm] (nn) {Stochastic\\neural network};

\matrix (outputs) [draw, fill=white, below of=nn, yshift=-0.5cm, matrix of nodes, row sep=0.5cm, column sep=0.5cm, nodes={draw, fill=gray!30, minimum width=1cm}] {
    $\hat{\mathbf{y}}_{i, 1}$ & $\hat{\mathbf{y}}_{i, 2}$ & \node[draw=none, fill=none] (dots) {$\cdots$}; & $\hat{\mathbf{y}}_{i, m}$ \\
};

\node[draw, fill=white, right of=outputs, xshift=4cm] (kde) {Kernel density\\estimator};
\node[draw, fill=gray!30, right of=kde, xshift=1.5cm] (prior) {$p(\hat{\mathbf{y}} | \mathbf{x}_i)$};

\node[draw, fill=white, right of=input, xshift=8cm] (nf) {Normalizing flow};
\node[draw, fill=gray!30, below of=nf, yshift=-0.5cm] (llh) {$p(\mathbf{y} | \hat{\mathbf{y}}) \approx p(\mathbf{y} | \hat{\mathbf{y}}, \mathbf{x}_i)$};

\node (times) [circle, draw, fill=white, below of=llh, yshift=-0.5cm] {$\times$};
\node (int) [circle, draw, fill=white, right of=times, xshift=0.5cm] {$\int d\hat{\mathbf{y}}$};
\node (post) [draw, fill=green!30, right of=int, xshift=1cm] {$p(\mathbf{y} | \mathbf{x}_i)$};

\draw[->] (input) -- (nn);
\foreach \i in {1, 2, 4} {
    \draw[->] (nn) -- (outputs-1-\i);
}
\draw[->] (outputs) -- (kde);
\draw[->] (kde) -- (prior);
\draw[->] (prior) -- (times);
\draw[->] (nf) -- (llh);
\draw[->] (llh) -- (times);
\draw[->] (times) -- (int);
\draw[->] (int) -- (post);
\end{tikzpicture}
}
\caption{The FCMLC posterior density estimation method presented in Sec.~\ref{subsec:fcmlc-overview}. Details of the stochastic neural network and normalizing flow architectures are specified in Fig.~\ref{fig:nn-architecture} and Fig.~\ref{fig:nf-architecture}, respectively.}
\label{fig:summary}
\end{figure*}
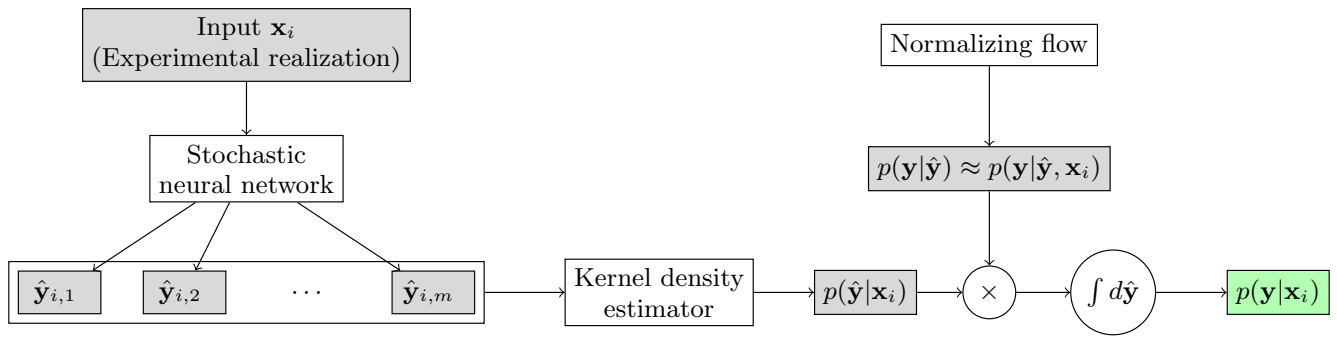

\subsection{\label{subsec:datagen}Data Generation}

Each experiment in this example study was previously fit using a frequentist framework with the \texttt{sblmc} code described in detail in Refs.~\cite{Diaz:2019fwt, Hardin:2022muu}. Performing global fits involves calculating the model log-likelihood (or equivalently, $\chi^2$) based on the experimental data. This calculation requires either a covariance matrix or a set of pull parameters, which are typically provided by the experiments in their data releases. The model expectations, combined with these cost functions, provide all the necessary information to generate a set of realizations for a given model (specified by the values of $U_{e4}$ and $\Delta m_{41}^2$) for a single experiment or a combination of experiments.

For experiments that provide a covariance matrix, realizations are constructed by adding the model expectation to residuals obtained by transforming the covariance matrix into a diagonal space, where independent Gaussian distributions can be sampled and then transformed back, thereby embedding any relevant correlation effects. In experiments that use pull parameters, realizations are generated by sampling these pull parameters (in a correlated fashion if specified) and then constructing the model for the sampled parameters.

A $101 \times 101$ grid even in log-space in $U_{e4}, \Delta m_{41}^2$ was defined, with $U_{e4} \in [0.001, 1 / \sqrt{2}]$ and $\Delta m_{41}^2 \in [0.01, 10.0] \,\text{eV}^2$. $50$ mock model realizations were computed for each of the experiments listed in Tab.~\ref{tab:experiments} for each of the $10,201$ combinations of the oscillation parameters to generate $510,050$ total samples. From here, training and test sets were delegated. Of the $10,201$ unique values of the oscillation parameters used to throw the Monte Carlo, a total of $10\%$ were withheld from the training set and dedicated exclusively to the test set (this 
subsample of the test set data is called the ``interpolation sample") to ensure that the model learns to regress in an abstract way. A further 1/9 of the remaining MC experimental results were then chosen at random to also join the test set (this subsample of the test set is referred to as the ``random sample"). This results in a total test set that is effectively $20\%$ of all available data. By splitting the test set into an interpolation and random sample, we can build a model which can best generalize and extrapolate model parameter predictions.

\subsection{\label{subsec:fcmlc-overview}The FCMLC Method for Posterior Density Estimation}

Given MC or observed experimental data $\mathbf{x}$, the first stage of building the posterior density over the oscillation parameters are dropout-at-inference NNs to directly predict $U_{e4}$ and $\Delta m_{41}^2$. We denote this prediction $\hat{\mathbf{y}}$. Networks used for the estimation of the underlying oscillation parameters were created and trained using Python's \texttt{tensorflow} API (Ref.~\cite{tensorflow2015-whitepaper}). The neural network architecture is shown in Fig.~\ref{fig:nn-architecture}. By calling this neural network $m$ times to predict the underlying oscillation parameters, $m$ independent samples of $\hat{\mathbf{y}}$ are made. By fitting a Kernel Density Estimate (KDE) on these $m$ samples, a probability distribution $p(\hat{\mathbf{y}} | \mathbf{x})$ can be generated. This KDE is fitted using the \texttt{statsmodels} Python package, Ref.~\cite{seabold2010statsmodels}.

It is desirable in the context of this problem to estimate the posterior probability distribution $p(\mathbf{y} | \mathbf{x})$, where $\mathbf{y}$ are the true underlying oscillation parameters that generated (via MC or otherwise) experimental data $\mathbf{x}$. This PDF can be computed by using the law of total probability:

\begin{equation}
p(\mathbf{y} | \mathbf{x}) = \int d \hat{\mathbf{y}} \, p(\mathbf{y} | \hat{\mathbf{y}}, \mathbf{x}) \, p(\hat{\mathbf{y}} | \mathbf{x}).
\label{eq:lotp}
\end{equation}

When considering raw experimental data, the first factor of the integrand of Eq.~\ref{eq:lotp} involves parameters with high dimension (in the case of STEREO, for instance, $\textrm{dim}\,\mathbf{x} = 126$), complicating density estimation a challenging problem. As such, we make the simplifying assumption that $p(\mathbf{y} | \hat{\mathbf{y}}, \mathbf{x}) \approx p(\mathbf{y} | \hat{\mathbf{y}})$. That is, $\mathbf{y}|\hat{\mathbf{y}}$ is approximately conditionally independent from $\mathbf{x}$, since the predictions $\hat{\mathbf{y}}$ from the well-trained first-stage NN have extracted all useful information from the experimental actualization $\mathbf{x}$. The quantity $p(\mathbf{y} | \hat{\mathbf{y}})$ can be estimated once by some sort of variational inference framework and used for all subsequent fits; in early stages of this work, a conditional multivariate kernel density estimator (KDE) was used, fit with endogenous entries $\mathbf{y}$ and exogenous entries $\hat{\mathbf{y}}$ (which are predictions from the neural network). We alter replaced the conditional KDE with a normalizing flow (NF) to handle conditional density estimation due to its generalizability and its significant reduction in computational resources. NFs were implemented using the the \texttt{normflows} package \cite{Stimper2023}. Details of the NF architecture are shown in Fig.~\ref{fig:nf-architecture}.

Eq.~\ref{eq:lotp} immediately makes apparent the procedure that can be used to compute $p(\mathbf{y} | \mathbf{x})$, which in turn can be used to construct credibility regions of the regression parameters $U_{e4}$ and $\Delta m_{41}^2$. The procedure described in this section is summarized in Fig.~\ref{fig:summary}. Though not a direct approximation of the Feldman-Cousins method, estimating model parameters from a posterior distribution is still an invaluable task for analyzing and understanding experimental data; because this procedure satisfies a similar role to the Feldman-Cousins method for understanding model fits, combined with its reliance on deep learning, we affectionately refer to this as ``Feldman-Cousins' ML Cousin" (FCMLC). 

Credibility regions are constructed numerically. To find the Bayesian credibility region at $1 - \alpha$ coverage, we numerically solve the equation below for $h_\alpha$:

\begin{equation}
    \int_{\mathbf{y} | p(\mathbf{y}|\mathbf{x}) \geq h_\alpha} d \mathbf{y} \, p(\mathbf{y} | \mathbf{x}) = 1 - \alpha
    \label{eq:confidence}
\end{equation}

\noindent The region of $\mathbf{y}$ space at $\alpha$ coverage, $\mathcal{Y}_\alpha$, is defined as $\mathcal{Y}_\alpha \equiv \{ \mathbf{y}\,|\,p(\mathbf{y} | \mathbf{x} ) \geq h_\alpha\} $. This method is used to construct the first-pass credibility regions shown in Fig.~\ref{fig:sample_fits}, and is especially useful for the analysis of signal-like data.

In the context of sterile neutrino fits, raster scans can also be used to interpret data consistent with null, as is done in NEOS \cite{neos:PhysRevLett.118.121802}. Once the conditional distribution for $\mathbf{y} | \mathbf{x} = (U_{e4}, \Delta m_{41}^2) | \mathbf{x}$ has been specified, the $1-\alpha$ confidence upper limit $U_{e4}^{\alpha}$ of the sterile mixing amplitude can also be found for for each value of $\Delta m_{41}^2$  by solving the equation:

\begin{equation}
    \int_{U_{e4}^{\alpha}}^{1/\sqrt{2}} p(U_{e4} | \Delta m_{41}^2, \mathbf{x})\,d U_{e4} = \alpha.
    \label{eq:raster}
\end{equation}

\noindent Comparisons of these two approaches are made in Ref.~\cite{lyons2014rasterscan2dapproach}; Ref.~\cite{feldman-cousins} points out that a strategy which does not require ``flip-flopping" between these two approaches is desirable. As such, we tend to favor Eq.~\ref{eq:confidence}, though both approaches may be encountered.

\subsection{\label{subsec:validation}Valuation Methods}

In an effort to convince the reader of the potential of FCMLC's utility for both single-experiment and global fits, we devise three parallel fitting tasks: fitting data from STEREO only (a $\overline{\nu}_e$ disappearance experiment that does not see a 3+1 signal, instead drawing an exclusion), BEST only (a $\nu_e$ disappearance experiment that sees a normalization effect consistent with a 3+1 signal), and a global fit of all experiments listed in Tab.~\ref{tab:experiments}. These three fits utilize data from fundamentally different experimental setups and each introduce their own challenges with data analysis, and are representative of some of the problems encountered in traditional global fits of disappearance experiments. Each of the steps listed below are performed on these three types of fits.

\subsubsection{Neural Network Performance}

To quantify the predictive performance of the fully-connected NNs predicted in this study, we report the training and test-set mean squared error (MSE), binned in each unique $U_{e4}, \Delta m_{41}^2$, and plotted in two dimensions with respect to the oscillation parameters. It is misleading to report the MSE for the entire training and test sets as single numbers; each experiment has regions of parameter space with that the experiment is not sensitive, and we do not expect the neural networks to extract the best-fit points in such circumstances, which would unfairly inflate the MSE.

\subsubsection{Coverage Checks}

Given that the Bayesian CRs we draw on the estimated posterior distribution are used for fit interpretation, it is important that these CRs are reasonable, recovering the true underlying model parameters the correct fraction of the time. The procedure used for performing these coverage checks is inspired by Ref.~\cite{Cook01092006}, which defines a correct method for verifying posterior CR coverage, requiring one to sample model parameters from the chosen prior distribution, and for each of these, sample data based on the likelihood function, and then compute the posterior distribution from which the coverage of an arbitrary CR can be inferred. Translating this procedure to the SBI approach, for 2000 trials, we generate a point in $(U_{e4}, \Delta m_{41}^2)$ space (uniformly in the logarithm of each parameter, according to the bounds specified in Sec.~\ref{subsec:datagen}), use \texttt{sblmc} to simulate experimental data from said model parameters, and then use \texttt{fcmlc} to build an estimate of the posterior distribution from which CRs can be drawn. We then compare observed versus expected coverages.

\subsubsection{Fits on Observed Experimental Data}

Once we have established that \texttt{fcmlc} gives reasonable posterior density distribution estimates, we use reported experimental data from the sources outlined in Tab.~\ref{tab:experiments} as inputs to the \texttt{fcmlc} procedure and report credibility regions in the model parameter space. We compare these with the published exclusion curve generated by STEREO \cite{Almazan2023}, the allowed region put forward by BEST \cite{PhysRevLett.128.232501} and, the allowed region (assuming Wilks') computed using the \texttt{sblmc} fitting framework \cite{Hardin:2022muu, Diaz:2019fwt}. Finally, we will use \texttt{fcmlc} to draw allowed regions for the oscillation parameters for a complete global fit of the $\nu_e$ disappearance data considered in this work.
\section{\label{sec:results}Results}

\begin{figure}[htp!]
    \centering
    
    \begin{subfigure}{\linewidth}
        \centering
        \includegraphics[width=0.95\linewidth]{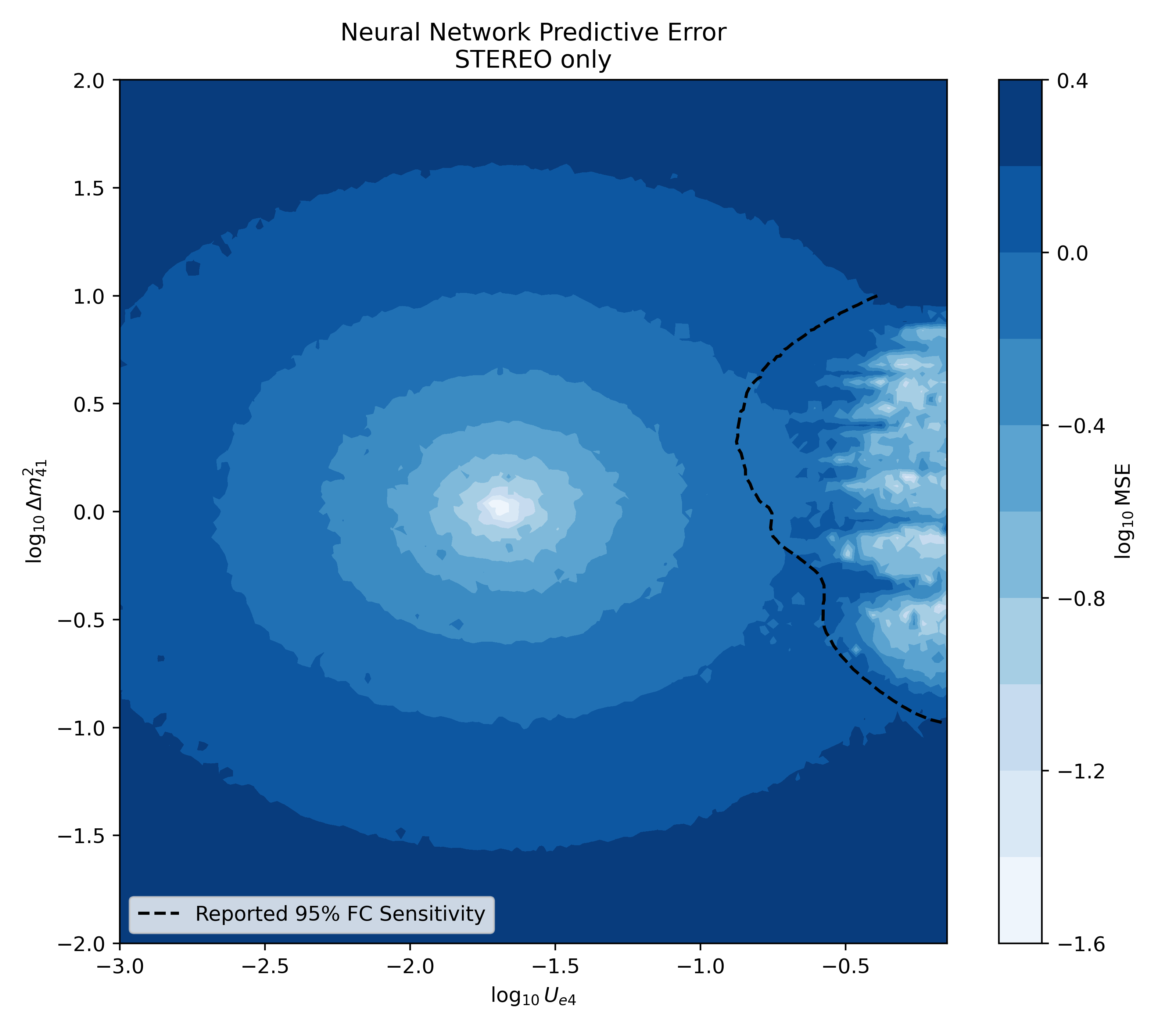}
    \end{subfigure}
    \vspace{-0.2cm}
    \begin{subfigure}{\linewidth}
        \centering
        \includegraphics[width=0.95\linewidth]{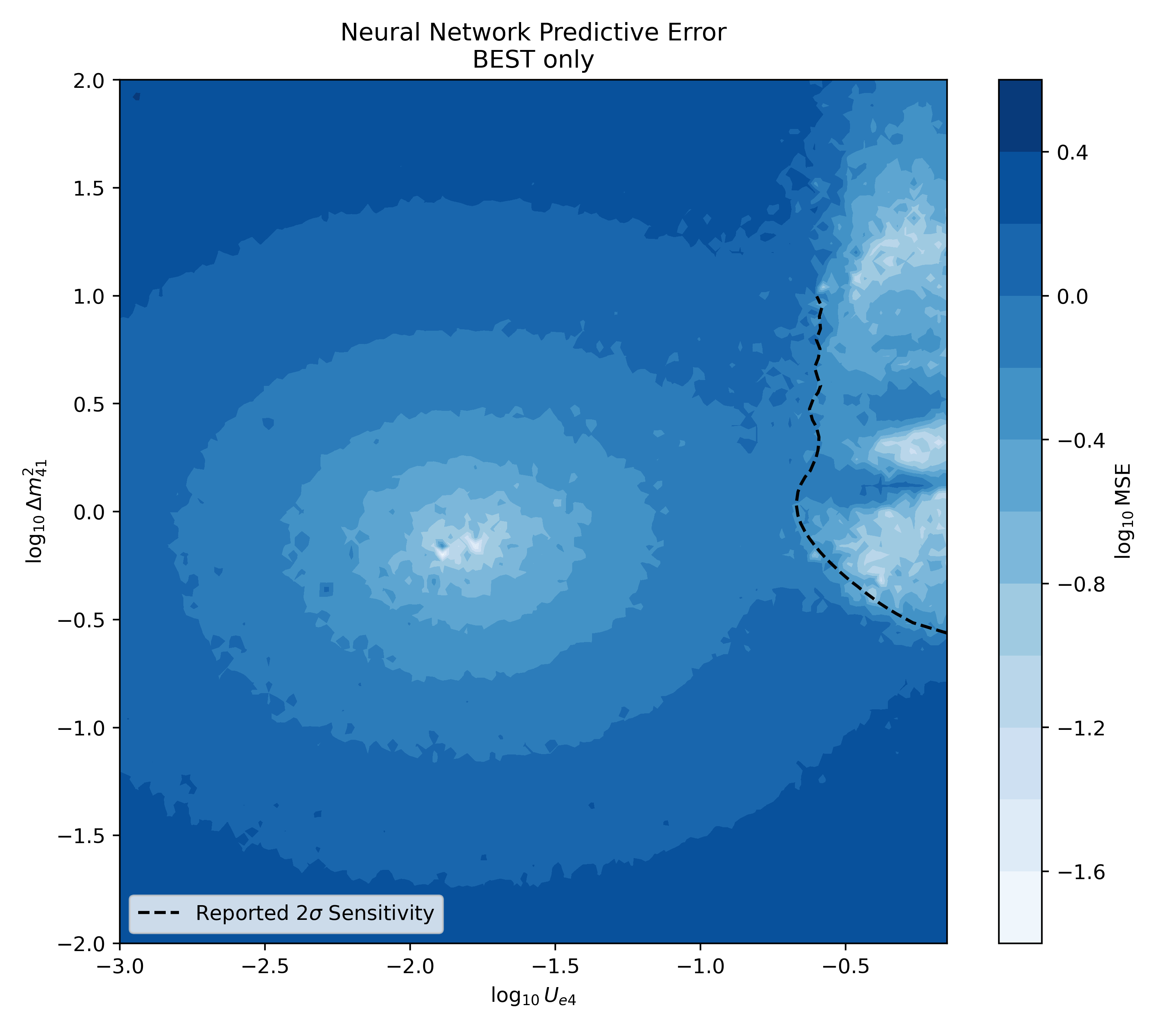}
    \end{subfigure}
    \vspace{-0.2cm}
    \begin{subfigure}{\linewidth}
        \centering
        \includegraphics[width=0.95\linewidth]{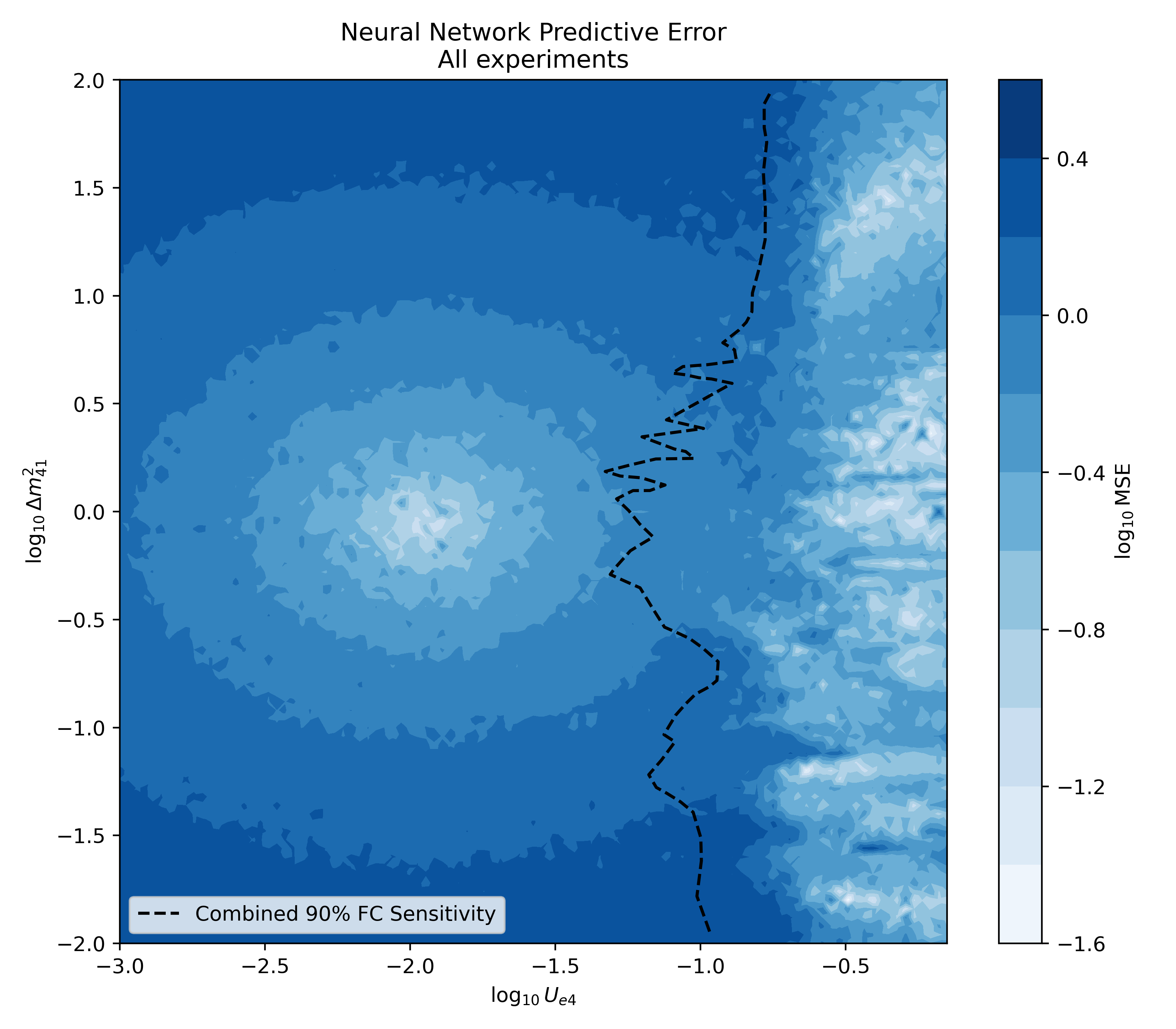}
    \end{subfigure}
    \caption{Binned neural network oscillation parameter test-set prediction accuracies as a function of true $U_{e4}, \Delta m_{41}^2$  for STEREO only (top), BEST only (middle), and a global fit (bottom).  See text for discussion of the circular pattern.  Published experimental sensitivities  are superimposed.}
    \label{fig:nn_performance}
\end{figure}

\subsection{Neural Network Predictive Performance}

The predictive accuracy of the neural networks trained in this study as a function of the true underlying oscillation parameters are shown in Fig.~\ref{fig:nn_performance}.

In this figure, important structures emerge: as hinted at earlier, neural networks are able to learn the true underlying oscillation parameters accurately only in certain regions of parameter space for each experiment (or combination of experiments for the global fit). One can compare the result to 
the published STEREO and BEST {\it sensitivities} \cite{Almazan2023, PhysRevLett.128.232501}, where the right of the line represents the region where a signal could be observed at a given confidence level. 
To obtain the global fit sensitivity, we run the Wilks'-based global fitting code described in Ref.~\cite{Diaz:2019fwt}.
In general, the regions where the networks learn well (white) corresponds well to the published sensitivities of each of the experiments. 

The circular contours seen in each of the plots in Fig.~\ref{fig:nn_performance} correspond to errors that are linear with respect to each of the oscillation parameters. This is because the neural networks, when presented with MC training data drawn from an underlying physical model to which the experiment(s) is (are) incapable of measuring reliably, are unable to find the best fit point; for the purposes of the loss optimization problem faced in NN training, the NN makes the $R^2 = 0$ guess of the average of these points in parameter space. This clear substructure in the prediction accuracy landscape is imperative for \texttt{fcmlc} to correctly draw exclusion curves when presented with a null model; it is how the algorithm ``learns" that it was shown data consistent with null.

\subsection{Example FCMLC Fits}

\begin{figure*}[hbtp!]
    \centering
    \begin{subfigure}[t]{0.4\textwidth}
        \centering
        \includegraphics[width=\textwidth]{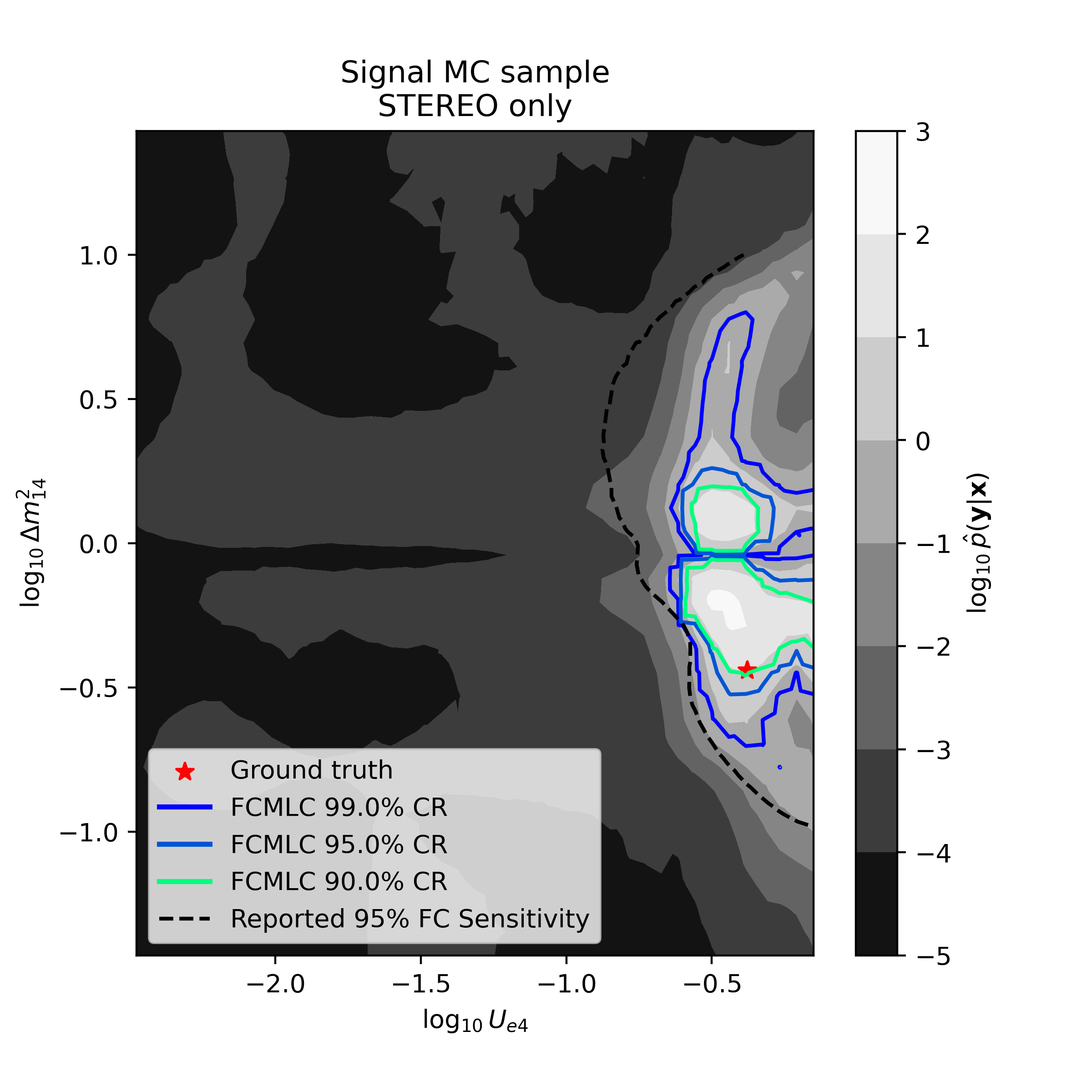}
    \end{subfigure}
    \hspace{0.02\textwidth}
    \begin{subfigure}[t]{0.4\textwidth}
        \centering
        \includegraphics[width=\textwidth]{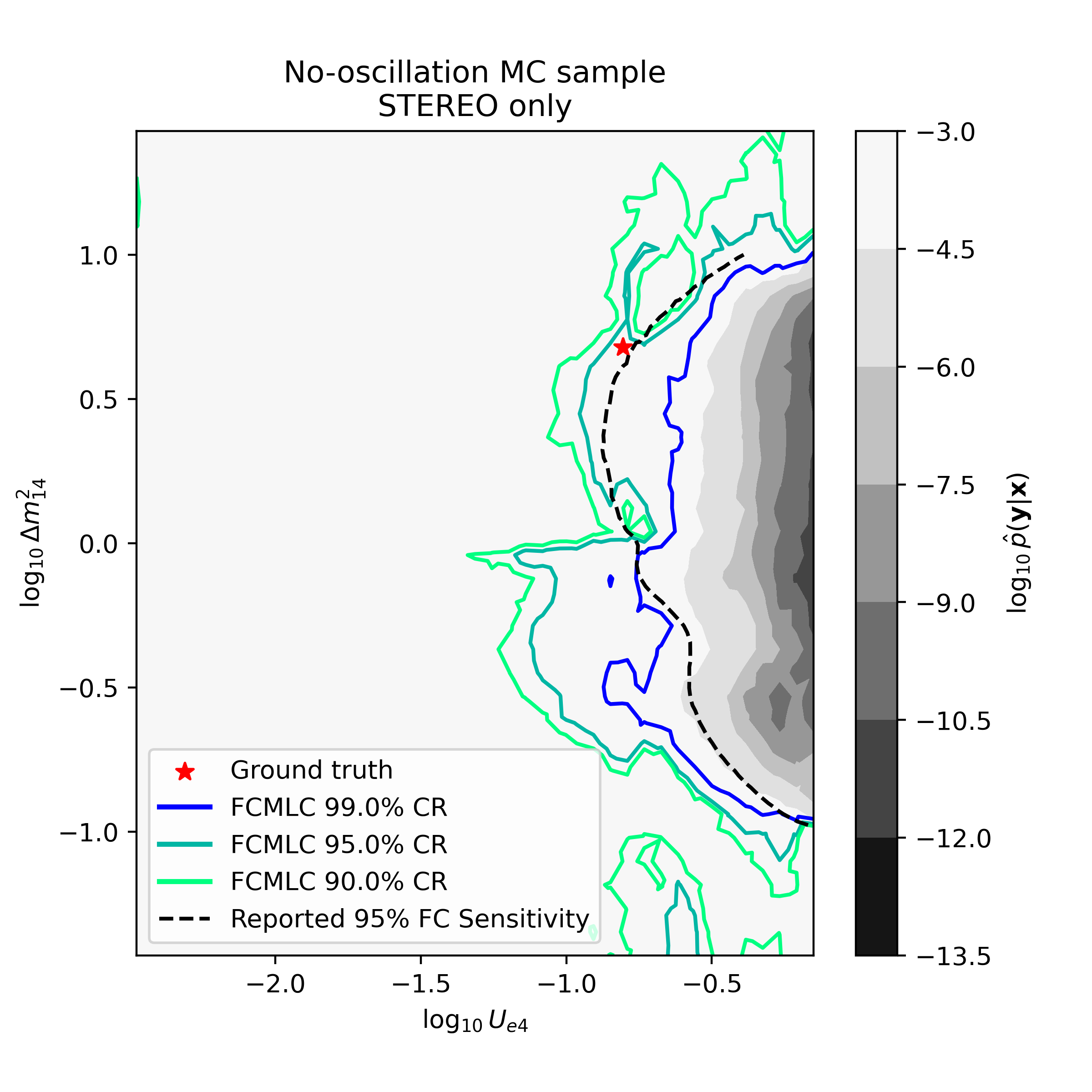}
    \end{subfigure}
    \vspace{-0.2cm}
    \begin{subfigure}[t]{0.4\textwidth}
        \centering
        \includegraphics[width=\textwidth]{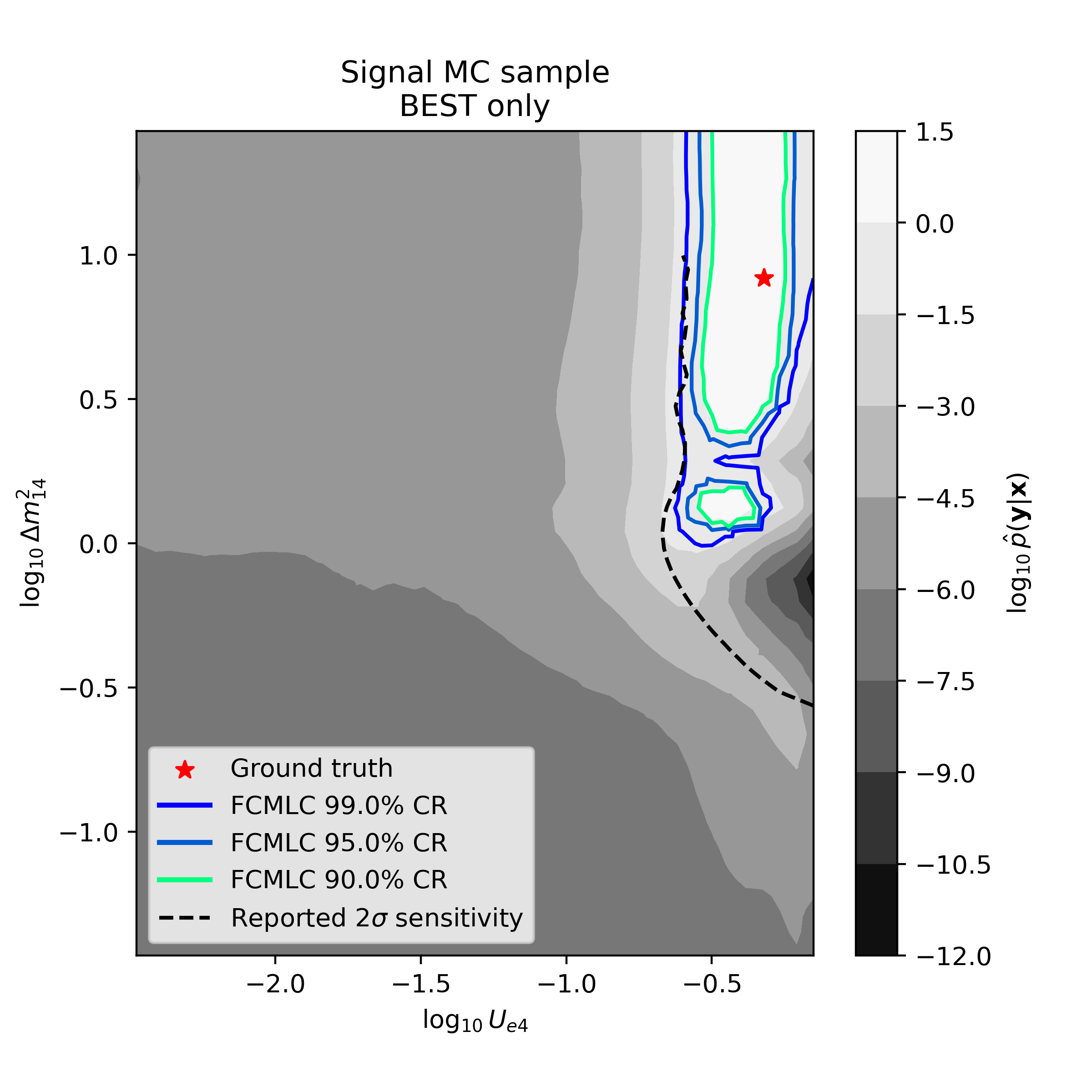}
    \end{subfigure}
    \hspace{0.02\textwidth}
    \begin{subfigure}[t]{0.4\textwidth}
        \centering
        \includegraphics[width=\textwidth]{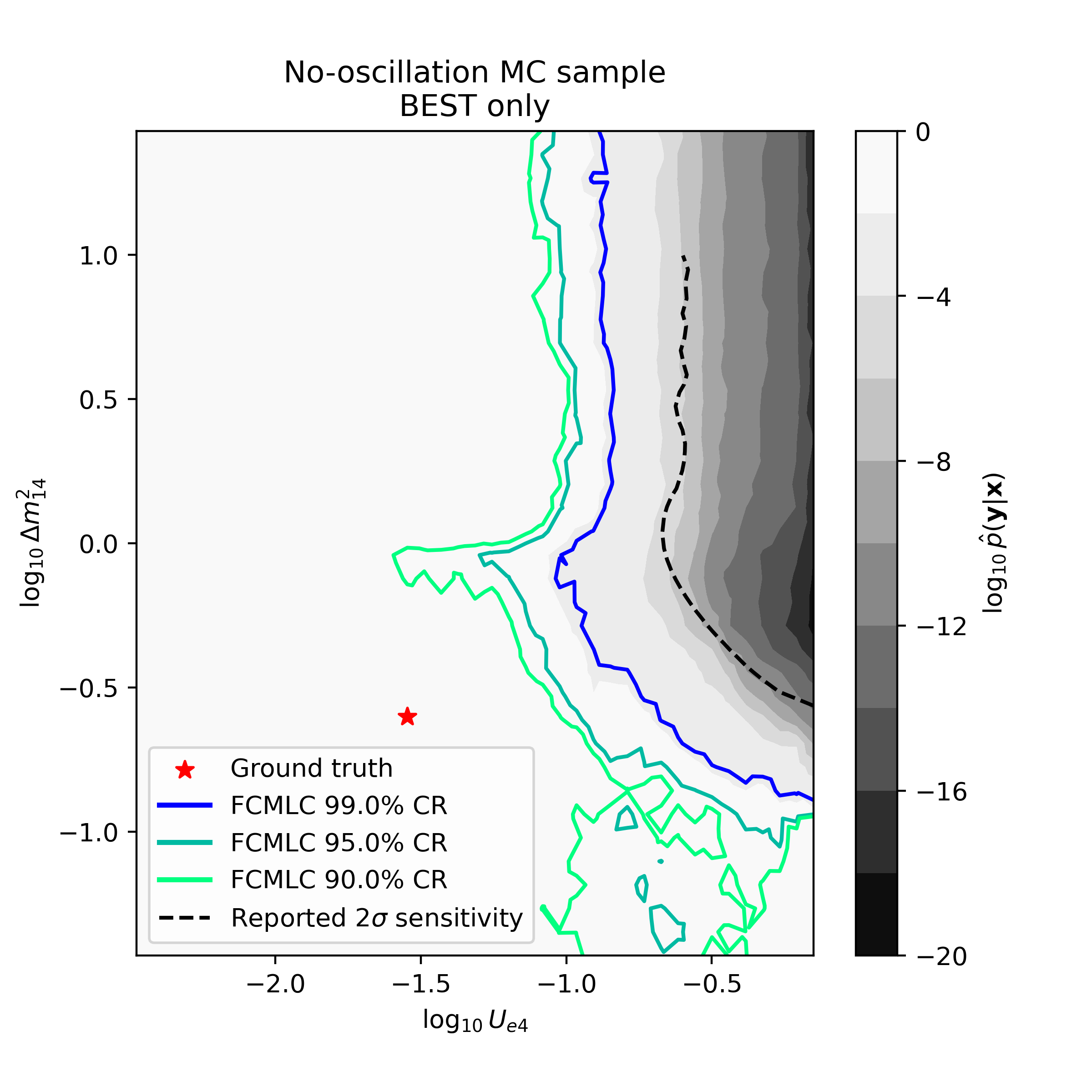}
    \end{subfigure}
    \vspace{-0.5cm}
    \begin{subfigure}[t]{0.4\textwidth}
        \centering
        \includegraphics[width=\textwidth]{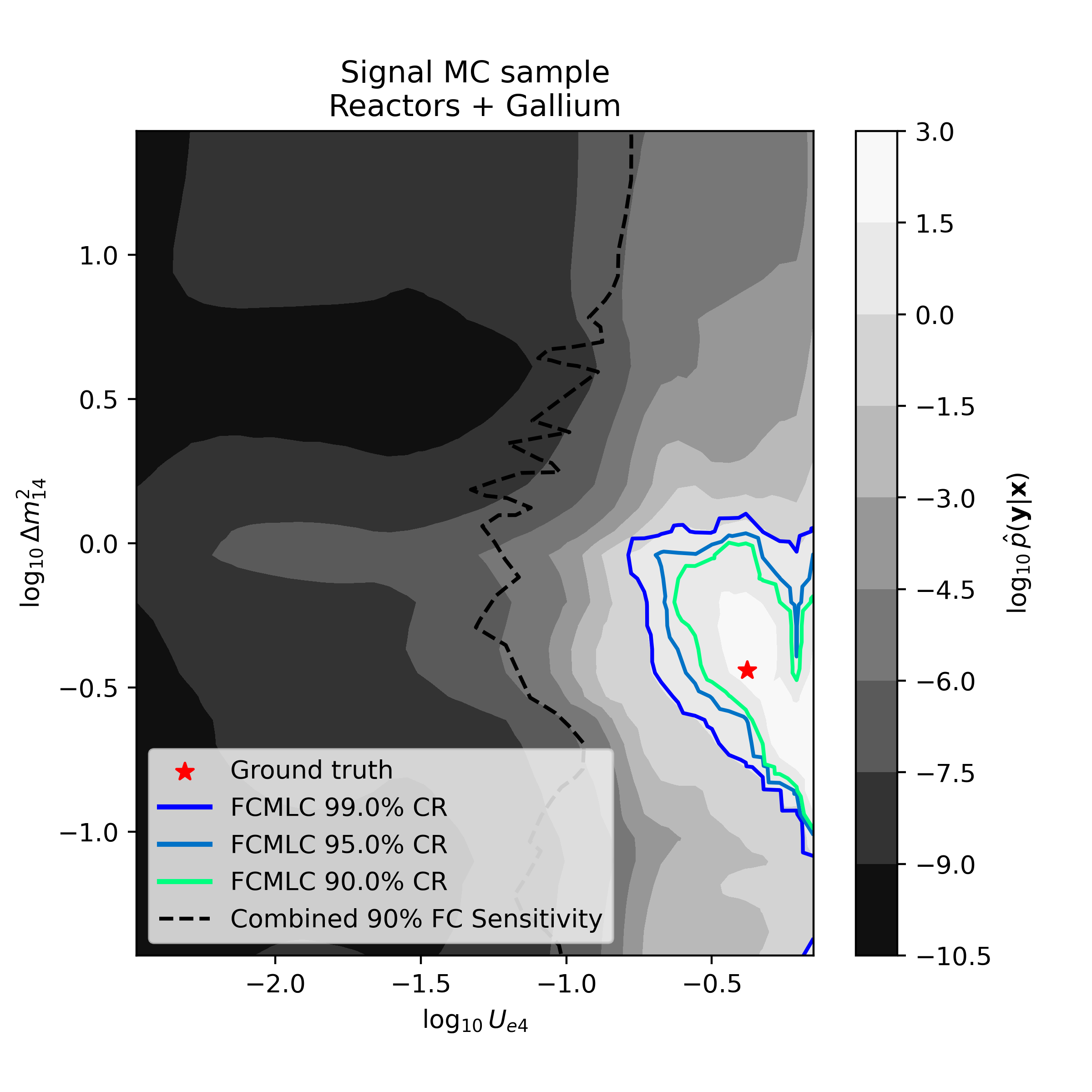}
    \end{subfigure}
    \hspace{0.02\textwidth}
    \begin{subfigure}[t]{0.4\textwidth}
        \centering
        \includegraphics[width=\textwidth]{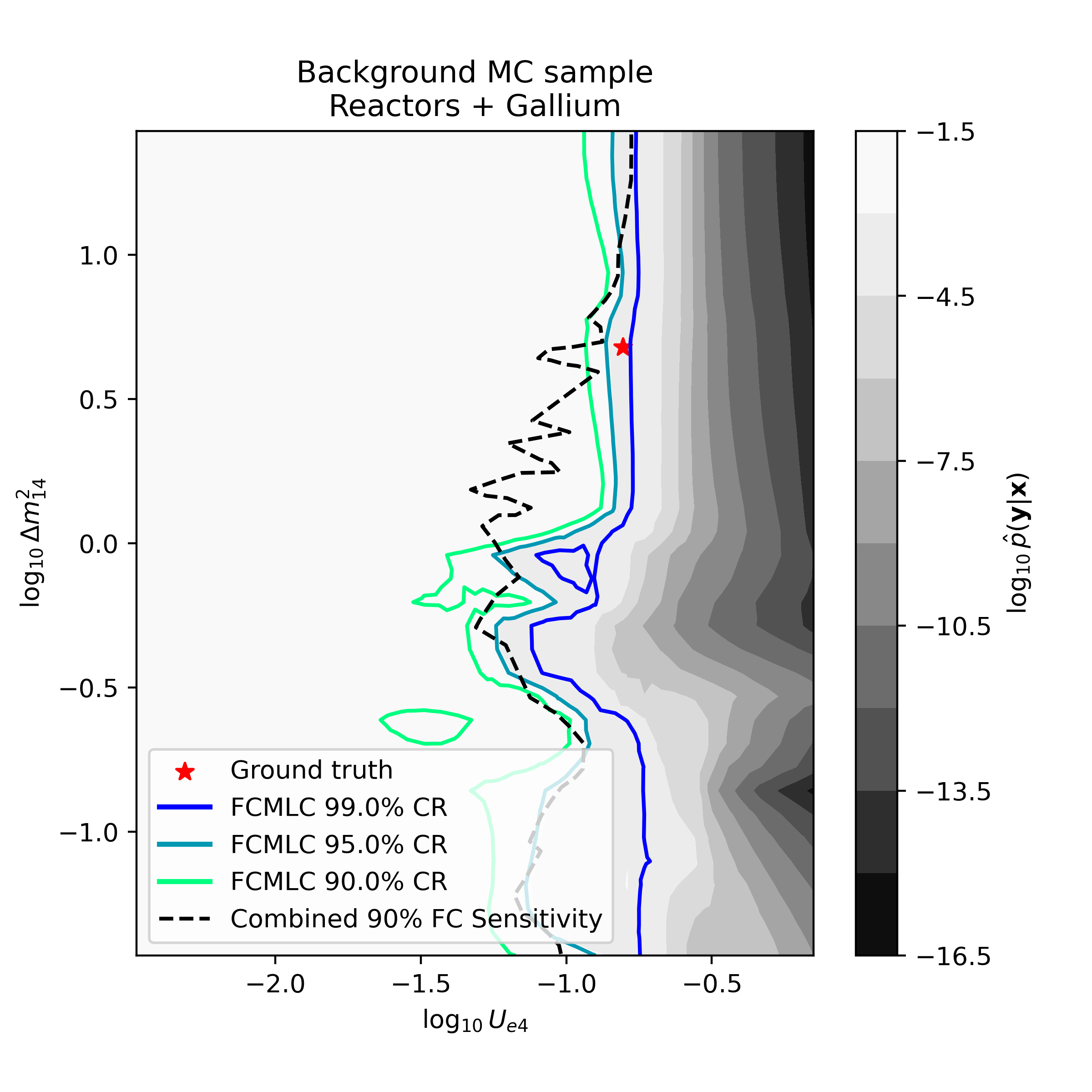}
    \end{subfigure}
    \caption{Sample \texttt{fcmlc} fits in the case of MC thrown from null-like and signal $3+1$ models. Reported sensitivities for the STEREO and BEST experiments are overlaid with the null-like fits to emphasize similarities in contour shape. Credibility regions were drawn according to the prescription in Eq.~\ref{eq:confidence}. Further tuning of \texttt{fcmlc} hyperparameters could improve smoothness of drawn CRs from the estimated posterior distribution, and will be explored in a future paper.}
    \label{fig:sample_fits}
\end{figure*}

Next, we explore how hypothetical signals would manifest for our three cases:  STEREO, BEST and the global fit.  In  Fig.~\ref{fig:sample_fits}, we present an array of sterile neutrino scenarios, where the underlying parameter points are indicated by red dots and labeled ``ground truth'').   In the cases where the ground truth is within the sensitivity of the experiment (to the right of the sensitivity line in Fig.~\ref{fig:nn_performance}), we expect the \texttt{fcmlc} to identify closed contours (that may extend off the plot at high $\Delta m^2$) around the red dot.   In the cases where the ground truth is outside of the computed sensitivity (to the left of the sensitivity line), the \texttt{fcmlc} fit not only draw open contours indicating that an allowed region is not identified, but contours which exclude regions of parameter space to which the experiment should be sensitive.

These plots illuminate striking abilities of the \texttt{fcmlc} approach. In the signal MC scenarios, the deep-learning procedure is able to produce closed contours around the ground truth oscillation model values. There is also remarkable consistency in the shapes of the posterior probability contours drawn by \texttt{fcmlc} with reported experimental sensitivities; a nearly identical shape to the sensitivity of BEST is reproduced by the \texttt{fcmlc} method on a null-like MC sample. Pitfalls of this method are discussed in Sec.~\ref{subsec:pitfalls}.

\subsection{Coverage Checks}
\begin{figure}[t!]
\centering
\includegraphics[width=0.8\linewidth]{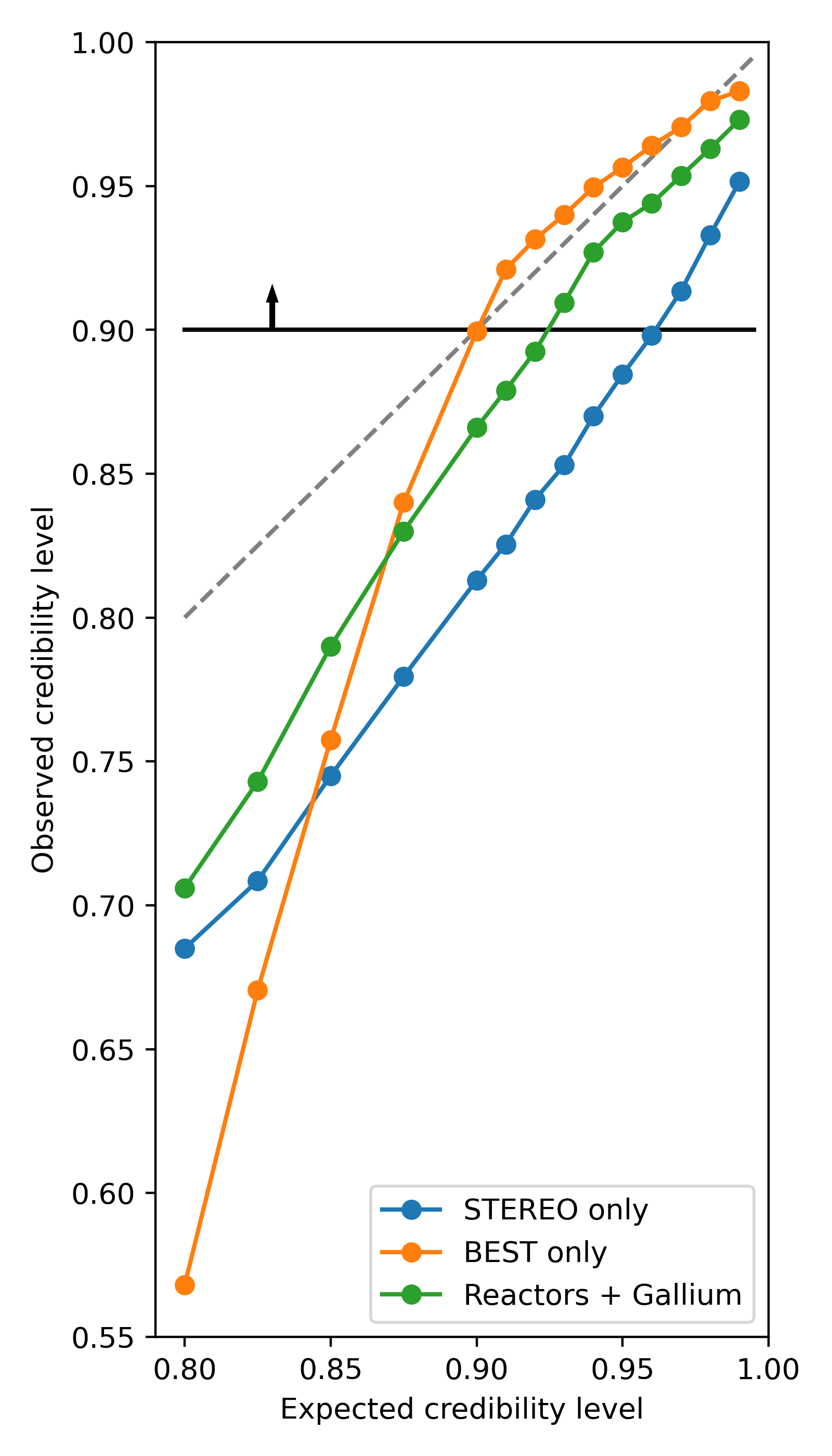}
\caption{Observed versus expected posterior CLs. A well-performing SBI procedure should show strong agreement between claimed and observed posterior coverages, lying near the gray dashed $45^\circ$ line. Though hyperparameter optimization strategies may be able to mitigate disagreement, we conjecture that the instability of low-credibility exclusion contours may be responsible for some of the mismatch seen in this figure. For further discussion, see Sec.~\ref{subsec:pitfalls}. Neutrino physicists are generally concerned with $>90\%$ CLs and CIs, so the region of space above the black horizontal line deserves the most attention.}
\label{fig:coverage}
\end{figure}

Through the procedure described in Sec.~\ref{subsec:validation}, we perform model parameter recovery checks and compute empirical and expected posterior CLs in Fig.~\ref{fig:coverage}. To reiterate, points in parameter space are sampled uniformly in log-space, matching the grid of parameter points used to generate training data. We address obvious limitations of this check in Sec.~\ref{subsec:pitfalls}.

\subsection{\label{subsec:realfits}Fits on Observed Experimental Data}

\begin{figure}[htp!]
    \centering
    \begin{subfigure}{\linewidth}
        \centering
        \includegraphics[width=0.83\linewidth]{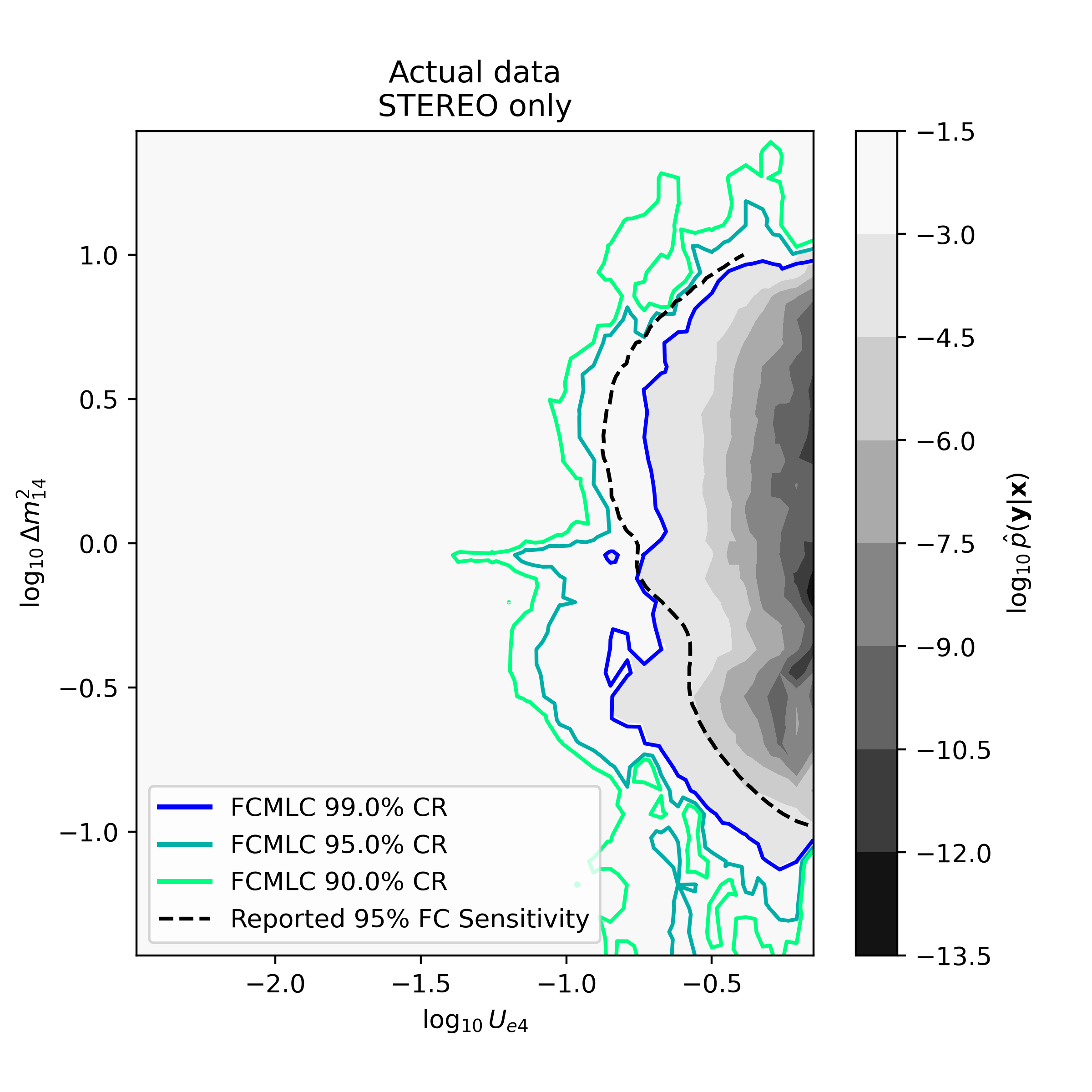}
    \end{subfigure}
    \vspace{-0.3cm}
    \begin{subfigure}{\linewidth}
        \centering
        \includegraphics[width=0.83\linewidth]{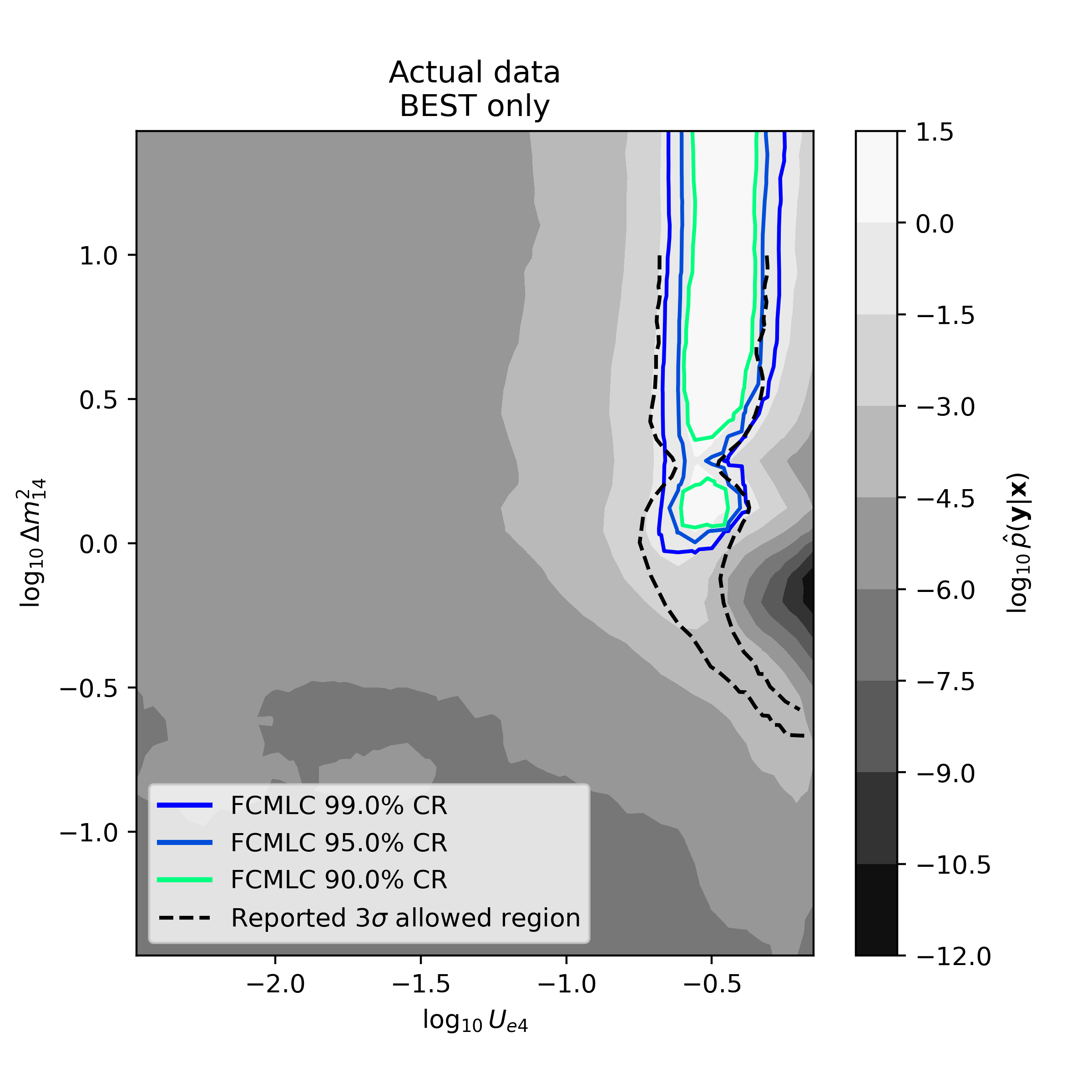}
    \end{subfigure}
    \vspace{-0.3cm}
    \begin{subfigure}{\linewidth}
        \centering
        \includegraphics[width=0.83\linewidth]{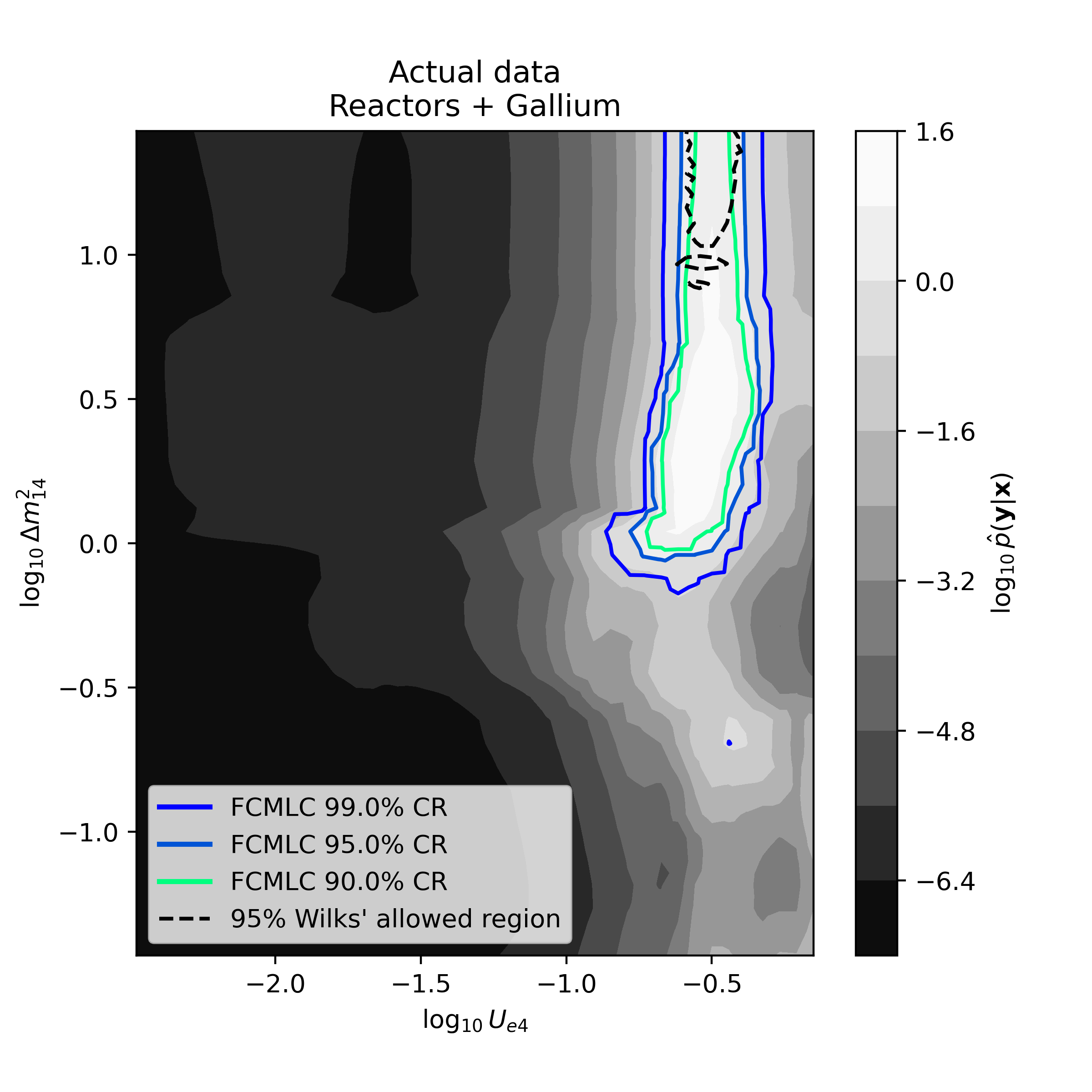}
    \end{subfigure}
    \caption{\texttt{fcmlc} fits on observed experimental data. For the STEREO-only fit (top), the colored exclusion curves corresponding to the \texttt{fcmlc} output were generated by the 2D credibility region construction (Eq.~\ref{eq:confidence}) to match the implementation in Ref.~\cite{stereocollaboration2020antineutrino}. Further tuning of \texttt{fcmlc} hyperparameters would smooth these drawn CRs.} 
    \label{fig:realfits}
\end{figure}

Having established reasonable performance using simulated experimental data, Fig.~\ref{fig:realfits} shows how
\texttt{fcmlc} interprets real-world experimental data from STEREO, BEST, and the $\nu_e/\overline{\nu}_e$-disappearance global fit. In the single-experiment cases, we compare the drawn exclusion curve to those drawn in the literature for STEREO \cite{Almazan2023} and the allowed-region published by BEST \cite{PhysRevLett.128.232501}.   We also compare the \texttt{fcmlc} result to a global fit result using the traditional Wilks'-based fit described in Ref. \cite{Diaz:2019fwt}.

Special attention should be paid to the fit results of BEST.  There is extraordinary topological agreement between the posterior distribution estimated by \texttt{fcmlc} and the $3\sigma$ exclusion curve reported by the collaboration.   We also observe reasonably good agreement between \texttt{fcmlc} and the STEREO exclusion. We see a qualitative match between the fit results from the SBI procedure and the regions expected under Wilks' theorem: both methods identify a localized high-credibility region in parameter space for $U_{e4} \approx 0.3$ ($\sin^2 2 \theta_{14} \approx 0.36)$, $\Delta m_{41}^2 \approx 10 \text{ eV}^2$, consistent with a normalization effect driven by the gallium experiments. In fact, the credibility regions drawn by \texttt{fcmlc} contain the traditionally computed best fit point at $(\sin^2 2 \theta_{14}, \Delta m_{41}^2) = (0.35, 9.3 \text{ eV}^2)$ \cite{Hardin:2022muu}. Deviations from Wilks' assumptions, such as non-Gaussianity in the likelihood or boundary effects in the parameter space drive certainly drive disagreement between the two methods, and further model tuning is likely to improve the SBI fit quality.

\section{\label{sec:discussion}Discussion}

\subsection{Timing Considerations}

Generation of the MC data described in Sec.\ref{subsec:datagen} completed in $\approx 2.4$K computing hours; these were expedited by running MC simulations in parallel. Neural networks and normalizing flows took on average $\approx 20$ minutes each to train on a CPU; we anticipate that the utilization of GPUs can make training faster. These two steps are by far the most resource intensive portion of this process, and are not yet fully optimized.

In contrast, the traditional Feldman-Cousins procedure requires both the MC generation above and the fitting of each realization.  Using the MCMC as implemented in the \texttt{sblmc} code, this takes approximately 2 hours per realization in this case.  The time taken for all fits is a total of $\approx 510$K computing hours,  $\approx 200$ times more than using FCMLC.  The time taken for these fits is dependent on the specific experiment and can vary by orders of magnitude in either direction.  Experiments such as BEST, with only 2 bins, and a simpler likelihood landscape, fit much faster than experiments such as STEREO, with more bins and more local minima.  

\begin{table}

\newcolumntype{P}[1]{>{\centering\arraybackslash}p{#1}}

\begin{tabular}{| *{4}{P{2cm}|}}  \hline

\multicolumn{4}{|c|}{\bf{Realization Generation}}  \\ \hline
\multicolumn{4}{|c|}{$2,400$ hrs}  \\ \hline \hline 
\multicolumn{4}{|c|}{\bf{Processing}}  \\ \hline

\multicolumn{2}{|c|}{\texttt{sblmc}} & \multicolumn{2}{c|}{\texttt{FCMLC}} \\ \hline

 Fit & $510,000$ hrs &  Normalizing Flows & $0.3$ hrs\\
\hline
 Feldman-Cousins & $\approx4$hrs &  Integration Evaluation & $0.1$ hrs\\
\hline
\hline
 Total & $\approx 510$khrs & Total  & $\approx 2.5$khrs\\
\hline
\end{tabular}
\caption {\label{tab:timing}  Comparison of the timing steps using \texttt{sblmc} (left) with a Feldman-Cousins procedure and the \texttt{FCMLC} (right) procedure.  Realization generation is shared between both processes and is not fully optimized, but is left in for comparison.} 
\end{table}

The integration procedure to combine the fitted distributions to the neural network predictions and the estimated conditional density from the normalizing flow via numerical integration consistently takes $\approx 5$ minutes; the integration over the intermediate two-dimensional $\hat{\mathbf{y}}$ (the predicted oscillation parameters from the neural networks) was completed on a $50 \times 50$ grid with evaluation of the posterior PDF on a $100 \times 100$ grid of true underlying model parameters $\mathbf{y}$. With MC generated, a neural network trained, and a normalizing flow trained, this is negligible compared to Feldman-Cousins; in fact, with MC generated, fits can be run in a Jupyter notebook. There is room for optimization in each of these routine steps; we suggest this as an avenue for further exploration in the case that the \texttt{fcmlc} method for posterior density estimation becomes more commonplace. 

\subsection{\label{subsec:pitfalls}Limitations}

The most obvious limitation of the \texttt{fcmlc} approach to fitting is its interpretability. \texttt{fcmlc} does not construct confidence levels for a rigorous statistical test, nor does it compute a posterior distribution given a well-defined, physically motivated prior over the model parameters. \texttt{fcmlc} sits in a class of its own; the posterior density that it draws is simply an AI's best guess for what model parameters could possibly fit the data seen. This does not mean, however, that \texttt{fcmlc} is not useful; this study at the very least shows that a prototypical \texttt{fcmlc} can reliably tell whether experimental data looks more signal- or background-like, with an associated level of uncertainty. Moreover, it could be used alongside higher-fidelity fitting frameworks like Feldman-Cousins to narrow relevant model parameter space to identify a best-fit point and a confidence interval, resulting in improved runtime.

The issue of imperfect credibility coverages has been a continuous source of criticism for a replacement of analytical fitting methods with SBI \cite{hermans2022a}; thus, the mismatch in coverages shown in Fig.~\ref{fig:coverage} is in good company.   Because neutrino physicists are generally
concerned with $>90\%$ CLs and CIs, where coverage is high, this issue is not crucial to the example global fits.  However, two issues related to the mismatch deserve discussion, and illustrate nuances that must be considered in a global fit.
First, experiments which see null-like results must draw exclusion curves, which in a Bayesian sense can be drawn from a contour of a posterior distribution, that in our case would look uniform in the case of a uniform prior. For an imperfect posterior approximation procedure such as \texttt{fcmlc}, reconstructing that distribution over a finite grid unavoidably introduces numerical fluctuations that negatively impact the recovery of injected parameters especially at lower credibility levels. Second, coverage mismatches are further complicated due to experimental idiosyncrasies when considering fits of parameters for a sinusoidal model like Eq.~\ref{PUe4}. Consider BEST: due to its poor detector segmentation and its low positional reconstruction resolution, it is substantially better at measuring a normalization rather than a shape-only effect due to sterile oscillations, and the credibility region shrinks to a vertical line in the $U_{e4}, \Delta m_{41}^2$ plane for which parameter recovery is made more difficult. 

This work is intended to present a proof-of-concept. As such, we have not spent time optimizing the architectures of the ML elements, tuning the KDE used to fit a PDF to the neural network predictions, testing different integration tools, or sampling data randomly rather than a grid. It is possible that with more careful calibrating, tighter and more centered contours around the ground truth model parameters can be drawn in the signal-like fits in Fig.~\ref{fig:sample_fits}, and more numerically stable and reported sensitivity-comparable contours can be drawn for the background-like fits. Although imperfect, the fact that \texttt{fcmlc} can produce fits that approximate the classic global fit result well with much shorter run-time, as seen in Tab.  \ref{tab:timing}  showcases its potential.

Anecodally, we noticed the way in which training data is sampled is dual to prior selection in a traditional Bayesian fit; here, we MC samples were generated from grid points uniformly separated in $\log_{10} U_{e4}$ and $\log_{10} \Delta m_{41}^2$ space. This matters in the context presented here, since exclusion curves drawn from an estimated posterior depend strongly on this choice.

\section{\label{sec:conclusion}Conclusion}

We anticipate that the \texttt{fcmlc} approach to fitting will be useful for fast analysis of experimental data, rapid evaluation of experimental sensitivities, and an important stage to initialize more localized, high-fidelity analysis frameworks such as the Feldman-Cousins algorithm. Before this happens, there are a few avenues for future work which we would like to identify:

\begin{itemize}
\item It is unknown how to best quantify the performance of this method against high-fidelity fits. Ideally, for many mock experimental realizations over a swath of underlying physical models, the output of \texttt{fcmlc} would be compared to fits performed with Feldman-Cousins. At the time of writing, this is a computationally impossible task. A method of reliably quantifying how well \texttt{fcmlc} approximates more sophisticated fitting frameworks is needed.
\item For a full-fledged global fit, each of the hyperparameters presented in the components of the framework should be optimized for the task presented. The fits to electron-neutrino disappearance data presented here were performed using ML component architectures that were ``good enough": a large-scale hyperparameter search is out of scope for this work and not necessary for the purposes of demonstrating the utility and reasonableness of this method.
\item Anecdotally, the reasonableness and the execution time of the fits from the \texttt{fcmlc} approach seem to be most bottlenecked by the density estimation procedures; namely, the KDE which fits a distribution over the predicted model parameters, and the conditional NF which relates the predicted model parameters to the true underlying ones. Such unsupervised learning problems are an area of active development; we suggest a review of these methods in the context of global fits which is also beyond the scope of this work.
\item In a sense, this method is plug-and-play. By this we mean each of the statistical learning components of the \texttt{fcmlc} approach to posterior density estimation (the dropout-at-inference NN, the KDE, the normalizing flow) can be swapped out for other probabilistic models. Further investigation into the right meta-architecture is warranted and again beyond the scope of this work.
\end{itemize}

This work highlights the critical need for fast and accurate global fitting methods in particle physics, particularly when traditional approaches are computationally prohibitive. By leveraging the power of deep learning, we have demonstrated a global fitting procedure having a significant reduction in runtime, offering a feasible prototypical alternative to the Feldman-Cousins method. We believe that the integration of machine learning techniques, as exemplified by our approach, will continue to play a transformative role in advancing the analysis capabilities in particle physics beyond sterile neutrino searches.

\begin{acknowledgments}
We are grateful to Phil Harris at MIT, and Gabriel Collin at the University of Adelaide. JV, JH and JMC thank MIT for support on this project. This material is based upon work supported in part by the National Science Foundation Graduate Research Fellowship under Grant No. 2141064.
\end{acknowledgments}

\bibliographystyle{iopart-num}
\bibliography{bibliography}

\onecolumngrid
\appendix
\newpage
\section{Network architectures}
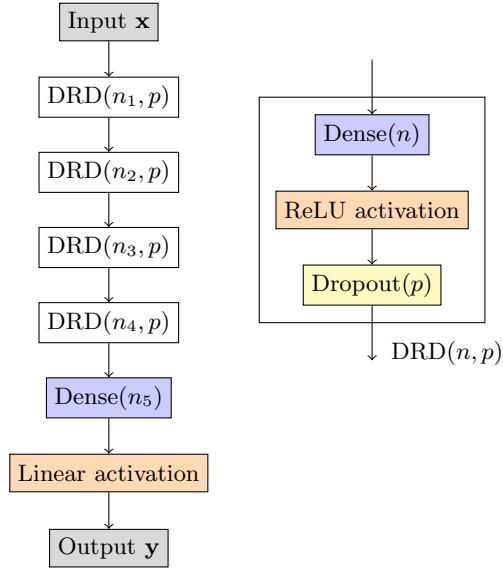
\begin{figure}[h!]
\centering
\begin{tikzpicture}[node distance=1.cm, every node/.style={rectangle, minimum width=.5cm, minimum height=0.5cm, draw, fill=blue!20, align=center}]

\node[draw, fill=gray!30] (input) {Input $\mathbf{x}$};
\node[draw, fill=white, below of=input] (drd1) {DRD($n_1, p$)};
\node[draw, fill=white, below of=drd1] (drd2) {DRD($n_2, p$)};
\node[draw, fill=white, below of=drd2] (drd3) {DRD($n_3, p$)};
\node[draw, fill=white, below of=drd3] (drd4) {DRD($n_4, p$)};

\draw[draw=black] (2.0,-4.0) rectangle ++(3.0,3.0);
\node[draw=none, fill=none, align=left] at (4.5,-4.4) {DRD($n, p$)};
\node (hiddenx) [right of=drd2, xshift=2.5cm, yshift=0.5cm] {Dense($n$)};
\node[draw, fill=orange!30, below of=hiddenx] (relux) {ReLU activation};
\node[draw, fill=yellow!30, below of=relux] (dropoutx) {Dropout($p$)};

\node (hiddenlast) [below of=drd4] {Dense($n_5$)};
\node[draw, fill=orange!30, below of=hiddenlast] (linac) {Linear activation};

\node[draw, fill=gray!30, below of=linac] (output) {Output $\mathbf{y}$};

\draw[->] (input) -- (drd1);
\draw[->] (drd1) -- (drd2);
\draw[->] (drd2) -- (drd3);
\draw[->] (drd3) -- (drd4); 
\draw[->] (drd4) -- (hiddenlast);
\draw[->] (hiddenlast) -- (linac);
\draw[->] (linac) -- (output);

\draw[->] (3.5, -0.5) -- (hiddenx);
\draw[->] (hiddenx) -- (relux);
\draw[->] (relux) -- (dropoutx);
\draw[->] (dropoutx) -- (3.5, -4.5);

\end{tikzpicture}
\caption{Architecture of the neural networks used for regression. As outlined in Sec.~\ref{subsec:nn-uncertainties}, the Dropout layers are left on at inference, which introduced sufficient stochasticity to perform the FCMLC algorithm. This network was compiled using mean squared error (MSE) as the loss, with an early stopping condition was set to ensure training convergence and prevent overfitting; training is halted if the $10\%$-split validation loss does not improve for 50 consecutive epochs. Additional information about the hyperparameters chosen for the test cases in this work are given in Tab.~\ref{tab:nn-hyperparams}.} 
\label{fig:nn-architecture}
\end{figure}

\begin{figure*}[t!]
\begin{center}
\resizebox{\linewidth}{!}{
\begin{tikzpicture}[node distance=1.cm, every node/.style={rectangle, minimum width=.5cm, minimum height=0.5cm, draw, fill=blue!20, align=center}]

\node[draw, fill=gray!30, xshift=0.5cm] (input) {Base distribution $\mathcal{N}_2 (\mathbf{0}, I_2)$};

\node[draw, fill=white, below of=input] (coupling1) {Coupling};
\node[draw, fill=pink, below of=coupling1] (lulp1) {LU linear permute};
\node[draw, fill=white, below of=lulp1] (coupling2) {Coupling};
\node[draw, fill=pink, below of=coupling2] (lulp2) {LU linear permute};
\node[draw, fill=white, below of=lulp2] (coupling3) {Coupling};
\node[draw, fill=pink, below of=coupling3] (lulp3) {LU linear permute};
\node[draw, fill=white, below of=lulp3] (coupling4) {Coupling};
\node[draw, fill=pink, below of=coupling4] (lulp4) {LU linear permute};

\draw[draw=black] (3.0,-7.0) rectangle ++(4.0,6.0);

\node[draw=none, fill=none, align=center] at (5,-.25) {$\mathbf{z}$};
\node[draw=none, fill=none, align=center] at (5,-7.75) {$\mathbf{z}'$};

\node[draw=none, fill=none, inner sep=0pt, minimum size=0pt] (A) at (4, -1.5) {};
\node[draw=none, fill=none, inner sep=0pt, minimum size=0pt] (B) at (6, -1.5) {};
\node[draw=none, fill=white, below of=A] (split1) {$\mathbf{z}_{1:d-1}$};
\node[draw=none, fill=white, below of=B] (split2) {$\mathbf{z}_{d:D}$};
\node[fill=white, below of=split1, right of = split1] (nn) {Sub-NN};
\node[draw=none, fill=white, below of=nn] (th) {$\bm{\theta}$};
\node[draw=none, fill=white, right of=th] (g) {$\mathbf{g}_{\bm{\theta}}$};
\node[draw=none, fill=white, below of=g] (split22) {$\mathbf{z}'_{d:D} := $\\$\mathbf{g}_{\bm{\theta}} (\mathbf{z}_{d:D})$};
\node[draw=none, fill=white, below of=split1, yshift=-2cm] (split12) {$\mathbf{z}'_{1:d-1} :=$\\$\mathbf{z}_{1:d-1}$};
\node[draw=none, fill=none, inner sep=0pt, minimum size=0pt] (A2) at (4, -6.5) {};
\node[draw=none, fill=none, inner sep=0pt, minimum size=0pt] (B2) at (6, -6.5) {};

\draw (5, -0.5) -- (5, -1.5);
\draw (A) -- (B);
\draw[->] (A) -- (split1);
\draw[->] (B) -- (split2);
\draw[->] (4, -3.5) -- (nn);
\draw[->] (nn) -- (th);
\draw[->] (th) -- (g);
\draw[->] (g) -- (split22);
\draw[->] (split2) -- (g);
\draw[->] (split1) -- (split12);
\draw (split12) -- (A2);
\draw (split22) -- (B2);
\draw (A2) -- (B2);
\draw[->] (5, -6.5) -- (5, -7.5);


\node[draw, fill=gray!30, below of=lulp4] (output) {Output distribution $\hat{p} (\mathbf{y} | \hat{\mathbf{y}})$};

\draw[->] (input) -- (coupling1);
\draw[->] (coupling1) -- (lulp1);
\draw[->] (lulp1) -- (coupling2);
\draw[->] (coupling2) -- (lulp2); 
\draw[->] (lulp2) -- (coupling3);
\draw[->] (coupling3) -- (lulp3);
\draw[->] (lulp3) -- (coupling4);
\draw[->] (coupling4) -- (lulp4);
\draw[->] (lulp4) -- (output);

\draw[draw=black] (7.5, -8.25) rectangle ++(4.0,6.5);
\node[draw=none, fill=none, right of=A, xshift=4.5cm, yshift=0.5cm] (intputsubnn) {$\mathbf{z}_{i:d-1}$};
\node[draw, below of=intputsubnn, yshift=-0.5cm] (hidden1) {Dense};
\node[draw, fill=white, below of=hidden1] (resblock1) {Residual block};
\node[draw, fill=white, below of=resblock1] (resblock2) {Residual block};
\node[draw, fill=white, below of=resblock2] (resblock3) {Residual block};
\node[draw, fill=white, below of=resblock3] (resblock4) {Residual block};

\draw[draw=black] (12.0, -9.5) rectangle ++(4.0,7.5);
\node (relua) [draw, fill=orange!30, right of=hidden1, xshift=4cm, yshift=-0.5cm] {ReLU activation};
\node (densea) [draw, below of=relua] {Dense};
\node (relub) [draw, fill=orange!30, below of=densea] {ReLU activation};
\node[draw, fill=yellow!30, below of=relub] (dropouta) {Dropout};
\node (denseb) [draw, below of=dropouta] {Dense};
\node (glu) [draw, fill=orange!30, below of=denseb] {GLU activation};

\node (plus) [circle, draw, fill=white, below of=glu, xshift=-1cm] {$+$};

\node (hiddenlast) [below of=resblock4] {Dense};

\node[draw=none, fill=none, below of=hiddenlast, yshift=-0.5cm] (outputsubnn) {$\bm{\theta}$};

\draw[->] (intputsubnn) -- (hidden1);
\draw[->] (hidden1) -- (resblock1);
\draw[->] (resblock1) -- (resblock2);
\draw[->] (resblock2) -- (resblock3);
\draw[->] (resblock3) -- (resblock4); 
\draw[->] (resblock4) -- (hiddenlast);
\draw[->] (hiddenlast) -- (outputsubnn);

\draw (13.5, -1.5) -- (13.5, -2.5);
\draw[->] (14.5, -2.5) -- (relua);
\draw[->] (relua) -- (densea);
\draw[->] (densea) -- (relub);
\draw[->] (relub) -- (dropouta);
\draw[->] (dropouta) -- (denseb);
\draw[->] (denseb) -- (glu);
\draw (glu) -- (14.5, -9);

\draw (14.5, -2.5) -- (12.5, -2.5);
\draw (12.5, -2.5) -- (12.5, -9);
\draw[->] (14.5, -9) -- (plus);
\draw[->] (12.5, -9) -- (plus);
\draw[->] (plus) -- (13.5, -10);

\node[draw=none, fill=none, align=center] at (0.5,1) {\textbf{Normalizing flow}};
\node[draw=none, fill=none, align=center] at (5,0.5) {\textbf{Coupling layer}};
\node[draw=none, fill=none, align=center] at (9.5,0.) {\textbf{Coupling layer sub-network}\\(Sub-NN)};
\node[draw=none, fill=none, align=center] at (14,-0.5) {\textbf{Residual block}};

\end{tikzpicture}
}
\end{center}
\caption{Architecture of the normalizing flows used for conditional density estimation. Coupling layers are based on smaller sub-neural networks that learn parameters for some sort of invertible transformation of intermediate variables; much research is devoted to the identification of useful parameters $\bm{\theta}$ and invertible transforms $\mathbf{g}_{\bm{\theta}}$. The normalizing flow was trained to minimize the Kullback-Leibler divergence $D_{KL}$ between training samples $\mathbf{y} | \hat{\mathbf{y}}$ and the estimated conditional probability $\hat{\mathbf{p}}(\mathbf{y} | \hat{\mathbf{y}})$. Additional information about the hyperparameters chosen for the test cases in this work are given in Tab.~\ref{tab:nf-hyperparams}.}
\label{fig:nf-architecture}
\end{figure*}
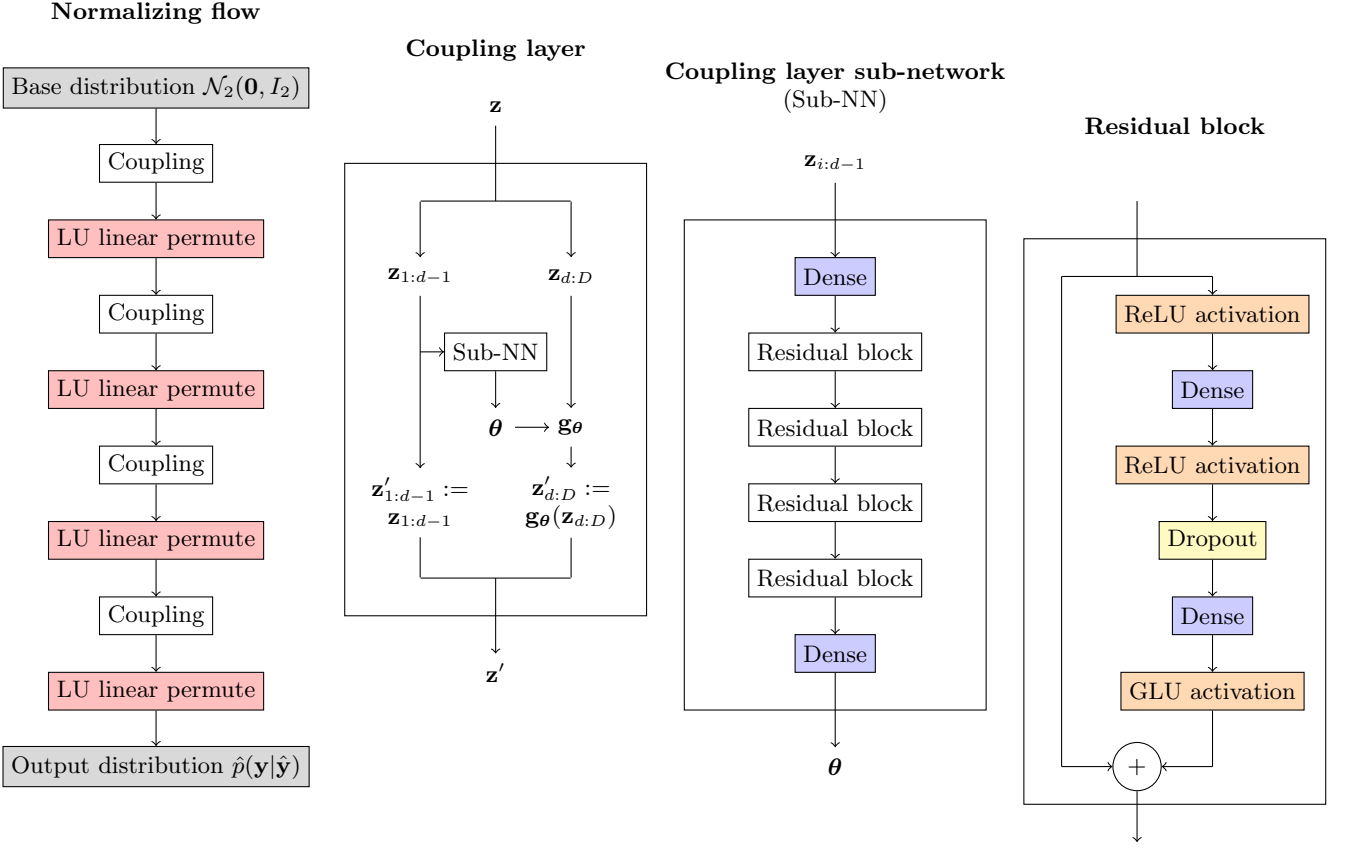

\newpage
\section{Hyperparameters for NN and NF architectures}
\begin{table*}[h!]
\centering
\begin{tabular}{|c|c|c|c|c|c||c|c|c|c|c|}  \hline
\multicolumn{6}{|c||}{Experiments} & \multicolumn{5}{c|}{Neural network hyperparameters} \\ \hline 
S & P & D & N & B & S/G & Layer sizes $(n_1,n_2,n_3,n_4,n_5)$ & Dropout rate & Optimizer & Learning rate & Batch size \\ \hline
\checkmark & & & & & & $(100,100,50,25,10)$ & $0.02$ & Adam & $0.001$ & $32$ \\ \hline
& & & & \checkmark & & $(100,100,50,25,10)$ & $0.02$ & Adam & $0.001$ & $32$ \\ \hline
\checkmark & \checkmark & \checkmark & \checkmark & \checkmark & \checkmark & $(150, 100, 50, 25, 25)$ & $0.02$ & Adam & $0.001$ & $32$ \\ \hline
\end{tabular}
\caption {\label{tab:nn-hyperparams}Neural network hyperparameters.}
\end{table*}

\begin{table*}[h!]
\centering
\begin{tabular}{|c|c|c|c|c|c||c|c|c|c|c|c|}  \hline
\multicolumn{6}{|c||}{Experiments} & \multicolumn{6}{c|}{Normalizing flow hyperparameters} \\ \hline 
S & P & D & N & B & S/G & Dropout rate & Training epochs & Optimizer & Learning rate & LR decay & Batch size \\ \hline
\checkmark & & & & & & $0.15$ & $3$ & Adam & $0.0005$ & $0$ & $512$ \\ \hline
& & & & \checkmark & & $0.10$ & $3$ & Adam & $0.0005$ & $10^{-5}$ & $512$ \\ \hline
\checkmark & \checkmark & \checkmark & \checkmark & \checkmark & \checkmark & $0.15$ & $3$ & Adam & $0.0005$ & $0$
 & $512$ \\ \hline 
\end{tabular}
\caption {\label{tab:nf-hyperparams}Normalizing flow hyperparameters.}
\end{table*}

\end{document}